\documentclass[review]{elsarticle}

\usepackage{lineno,hyperref}
\usepackage{graphicx}
\usepackage{comment}
\usepackage{amssymb}
\usepackage{lineno}
\usepackage{lineno,hyperref}
\usepackage{amsmath,amsfonts,amsthm,amssymb,graphicx,subfigure,mathtools,tikz,hyperref,bm,blindtext,float}
\usepackage{diagbox}
\usepackage{multirow}
\usepackage{geometry}
\usepackage{algorithm}
\usepackage{algpseudocode}

\modulolinenumbers[5]

\journal{Journal of \LaTeX\ Templates}

\begin{document}

\begin{frontmatter}

\title{A fast multi-fidelity method with uncertainty quantification for complex data correlations:
Application to vortex-induced vibrations of marine risers}

\author[brown]{Xuhui Meng}
\author[brown]{Zhicheng Wang}
\author[mit,sea]{Dixia Fan}
\author[mit,sea]{Michael Triantafyllou}
\author[brown]{George Em Karniadakis \fnref{2}}

\address[brown]{Division of Applied Mathematics, Brown University, Providence, RI 02906, USA}
\address[mit]{Department of Mechanical Engineering, Massachusetts Institute Technology, Cambridge, MA 02139, USA}
\address[sea]{Sea Grant College Program, Massachusetts Institute
of Technology, Cambridge, MA 02139, USA}
\fntext[2]{Corresponding author: george\_karniadakis@brown.edu}



\begin{abstract}
We develop a fast multi-fidelity modeling method for very complex correlations between high- and low-fidelity data by working in modal space to extract the proper correlation function. We apply this method to infer the amplitude of motion of a flexible marine riser in cross-flow, subject to vortex-induced vibrations (VIV).  VIV are driven by an absolute instability in the flow, which imposes a frequency (Strouhal) law that requires a matching with the impedance of the structure; this matching is easily achieved because of the rapid parametric variation of the added mass force.  As a result, the wavenumber of the riser spatial response is within narrow bands of uncertainty.  Hence, an error in wavenumber prediction can cause significant phase-related errors in the shape of the amplitude of response along the riser, rendering correlation between low- and high-fidelity data very complex.  Working in modal space as outlined herein, dense data from low-fidelity data, provided by the semi-empirical computer code VIVA, can correlate in modal space with few high-fidelity data, obtained from experiments or fully-resolved CFD simulations, to correct both phase and amplitude and provide predictions that agree very well overall with the correct shape of the amplitude response. We also quantify the uncertainty in the prediction using Bayesian modeling and exploit this uncertainty to formulate an active learning strategy for the best possible location of the sensors providing the high fidelity measurements.
\end{abstract}

\begin{keyword}
\texttt{elsarticle.cls}\sep \LaTeX\sep Elsevier \sep template
\MSC[2010] 00-01\sep  99-00
modal decomposition \sep multi-fidelity \sep uncertainty quantification \sep active learning \sep vortex-induced vibrations
\end{keyword}

\end{frontmatter}


\section{Introduction}

\begin{figure}[!htb]
    \centering
    \includegraphics[width=0.95\textwidth,trim=2 2 2 2,clip]{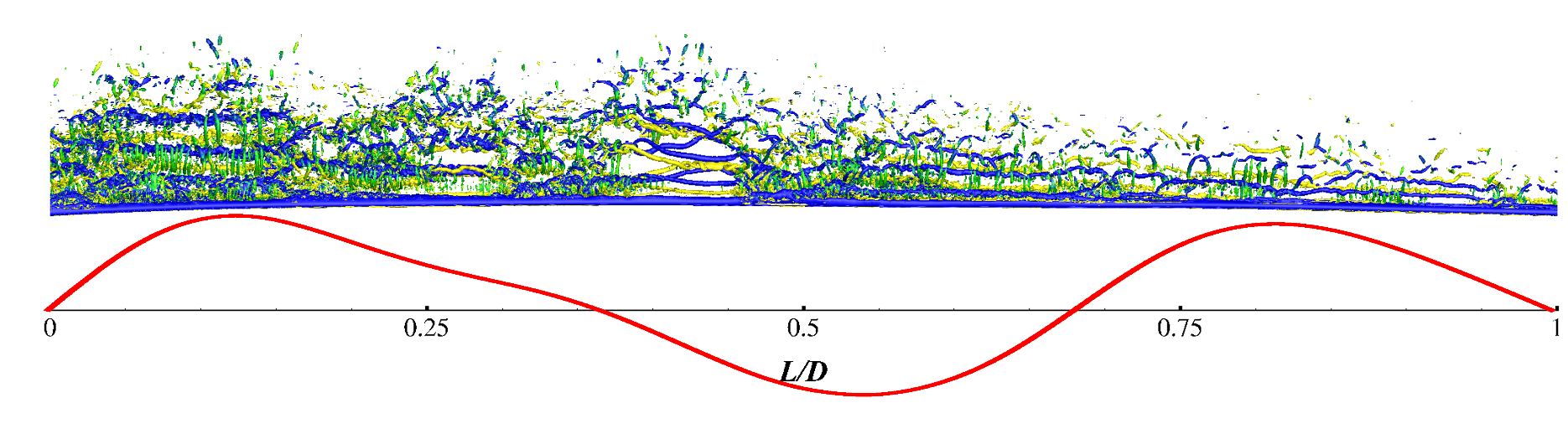}
    \caption{Large eddy simulation of the vortex induced vibration of a flexible riser in linearly sheared flow. Top figure, instantaneous vortical flow patterns; bottom figure, the corresponding riser displacement (red curve). Note that, the magnitude of the displacement has been enlarged in order to visualize it more clearly.}
    \label{fig:viv_les}
\end{figure}

{\bf Vortex-Induced Vibrations.} Long slender structures within steady oncoming flow are subject to vibrations caused by vortical structures forming due to a distributed flow instability in their wake \cite{wu2012review}, as shown in Fig. \ref{fig:viv_les}, and the flexible cylinder deflects due to an increase in drag caused by vortex shedding. The problem has considerable theoretical interest as it constitutes a fundamental nonlinear flow-structure interaction (FSI) system \cite{williamson2004vortex}, while it is very important for the design of offshore industry systems, such as risers, cables, and hawsers, which are subject to large drag loads and potentially catastrophic fatigue damage as a consequence of the vortex-induced vibration (VIV) \cite{wang2020review}. Therefore, a large amount of research has focused on better understanding and predicting the flexible cylinder VIV response, as well as devising suppression methods to mitigate excessive fatigue damage. 

Similar to rigid cylinder VIV \cite{gabbai2005overview,williamson2008brief, bearman2011circular}, flexible cylinder VIV display FSI features, including ``lock-in" \cite{bearman1984vortex}, dual resonance of in-line (IL) and cross-flow (CF) vibration \cite{dahl2007resonant,FanVIVJFM2019}, fluid-structure energy exchange that is high correlated with the form of the IL-and-CF motion trajectory \cite{dahl2006two,bourguet2011wake,fan2017vortex}, and large variation of the effective fluid added mass at different flow speeds \cite{williamson1996vortex, sarpkaya2004critical}. However, flexible cylinder VIV have additional features caused by the distributed flow instability along their length, which can cause multiple frequencies of response and variable spatial correlation coherence of the vortical patterns. A few well-controlled laboratory experiments and large-scale field test have helped to shed some light on these features. Fan et al. \cite{FanVIVJFM2019} measured directly the displacement of a flexible cylinder in uniform flow over a wide range of reduced velocities at moderate Reynolds number using an underwater optical measurement system. Their work showed that various modes are excited in the IL and the CF directions, documented jumps in maximum amplitude as function of oncoming flow velocity, and mapped the distributed hydrodynamic coefficient changes during a modal switch. Similar results were reported by Chaplin et al. \cite{chaplin2005laboratory} from a larger-scale experiment of a flexible cylinder in a stepped incoming flow. The results were compared with predictions from various numerical and semi-empirical simulations \cite{chaplin2005blind}. The predictions, although capturing some flexible cylinder VIV features, also showed discrepancies from the experiments. Under the Norwegian Deepwater Program (NDP), experiments were conducted on a 38-m flexible riser model in both uniform and linearly sheared flow \cite{braaten2004ndp}. It was shown that the riser model would exhibit chaotic response, switching randomly between steady-state and chaotic states, with strong high harmonic components in the lift force during steady-state response \cite{modarres2010effect,modarres2011chaotic,zheng2014fatigue}. Field tests, such as towing long flexible cylinders within the Gulf Stream \cite{vandiver2005high,vandiver2006fatigue,vandiver2009insights}, displayed also the presence of mixed traveling and standing waves, a strong interaction between IL and CF modal responses, and a response that consisted of multiple frequencies.

Various computational fluid dynamics (CFD) tools have been developed to predict the structural and wake response of flexible cylinders in cross-flow \cite{lucor2001vortex, bourguet2011vortex, bourguet2013distributed,zhu2018wake}, but due to the large computation resources required, CFD research has focused on relatively low Reynolds numbers ($Re \sim O(10) - O(10^3)$) and small to medium aspect ratios ($L/d \sim O(10) - O(10^3)$) , which are different from full-scale conditions in riser deployed in offshore platforms ($Re \sim O(10^5)$ and $L/d \sim O(10^4)$). Therefore, the state-of-the-art riser VIV prediction tools widely adopted in the industry are often based on semi-empirical model predictions \cite{triantafyllou1999pragmatic, larsen2001vivana, roveri2001slenderex}. These codes have been developed based on the assumption of strip theory \cite{FanVIVJFM2019,wang2020large} and employing various spatial correlation models. The hydrodynamic model is hence simplified, adapting the hydrodynamic coefficients of a rigid cylinder conducting forced vibrations within a cross-stream \cite{gopalkrishnan1993vortex,fan2019robotic}. A key factor for riser VIV prediction via the semi-empirical approach is the availability of accurately measured hydrodynamic coefficients as distributed along the riser span. However, various studies showed that these coefficients are sensitive to several parameters, including Reynolds number \cite{xu2013experimental}, riser configuration \cite{chen2013hydrodynamic}, and surface roughness \cite{chang2011viv}. Hence, a systematic development of a single hydrodynamic database is virtually impossible. In addition, during the lifetime of a riser in the field, long-term effects, such as equipment aging and biofouling, inevitably alter the hydrodynamic coefficients, making long-term riser prediction and monitoring even more challenging. 

The objective of this study is to develop a fast vibration inference system by integrating low fidelity, low cost, semi-empirical prediction tools with high fidelity, high cost numerical and experimental approaches. Here, the high fidelity data consists of three sources: First, we use LES simulation results of VIV of a flexible cylinder in linearly sheared inflow. In the simulation, the reduced velocity is $U_r=\frac{U_m}{f_{n1}D}=15.65$, where $U_m=(U_{max}+U_{min})/2$ is the mean inflow velocity, $f_{n1}$ is the first modal natural frequency, $D$ is diameter of the cylinder. Here $U_{max}=1.4$ and $U_{min}=0.6$ are the inflow velocity at two ends of the current profile, respectively. The Reynolds number $Re=U_m\,D/\nu=800$, where $\nu$ is the kinematic viscosity. Moreover, in the simulation, the aspect ratio of the flexible cylinder $L/D=240$. Second, we use optical measurement experiments that were conducted at the MIT Tow Tank facility, using a flexible cylinder with $L/D=240$, towed to generate a uniform inflow at various speeds, to achieve $Re$ from 250 to 2,300, while the $U_r$ varies from 4.8 to 36. Note that, most of the high fidelity data presented in this paper are from the optical experiment. Third, we use the measurement data from NDP case 2430 \cite{braaten2004ndp}, to demonstrate our method is capable of predicting the VIV motions accurately in realistic conditions. In the NDP experiment a bare riser of $L=38$ m and $D=0.027$ m is towed horizontally at $U_{min}=0$, $U_{max}=1.5$ m/s. Having the high fidelity data, we subsequently performed calculations using the VIVA program \cite{triantafyllou1999pragmatic} to obtain the low fidelity data, with the same flow and structural parameters.

{\bf Multi-fidelity Modeling.} To predict the VIV response of a riser, we combine a dense set of data from a low fidelity model because of its low cost of use, with a relatively sparse set of experimental and field measurements that are expensive to obtain. With abundant simulation data from VIVA and a small set of accurate measurements, a natural idea is to utilize multi-fidelity modeling, such as multi-fidelity Gaussian process regression (GPR) \cite{forrester2007multi,perdikaris2017nonlinear,bonfiglio2018multi,costabal2019multi,tian2020enhanced}, or multi-fidelity deep neural networks \cite{meng2020composite,zhang373multi}, to assimilate data with different accuracy. In particular, we will use here the VIVA data as low-fidelity and experimental measurements as high-fidelity data. 

The key idea of multi-fidelity modeling is to discover the cross-correlation between the low- and high-fidelity data rather than approximating the high-fidelity data directly. Conventional multi-fidelity modeling approaches work well for cases in which the correlation is easier to obtain than the high-fidelity function itself. However, the correlation between the low- and the high-fidelity data in the present study is much more complicated than the high-fidelity function, due to phase errors between the multi-fidelity data. 
The aforementioned multi-fidelity GPR \cite{perdikaris2017nonlinear} has been extended to resolve the complicated correlation caused by the phase error between the low- and high-fidelity data by adding more dimensions, i.e., shifts of low-fidelity functions,  to the input space. To obtain accurate predictions, additional dimensions are used for various low- and high-fidelity data in \cite{lee2019linking}, but how to choose a proper number of additional dimensions remains unclear. Furthermore, the increasing input dimensions will clearly lead to higher computational cost.
Therefore, we need to develop alternative multi-fidelity approaches, which can overcome these drawbacks.
To this end, we develop a multi-fidelity approach, which we refer to as VIV-MFnet, to assimilate the VIV simulation and the experimental data that can handle phase errors effectively and efficiently. The rest of the paper is organized as follows: we present the VIV-MFnet method in Sec. \ref{sec:method}, and we show our results in Sec. \ref{sec:results}. Finally, we give a summary for this work in Sec. \ref{sec:summary}.

\section{Methodology}
\label{sec:method}

\subsection{Fourier decomposition of displacement}


\begin{figure}
\centering
\subfigure[]{\label{fig:examplea}
\includegraphics[width=0.45\textwidth]{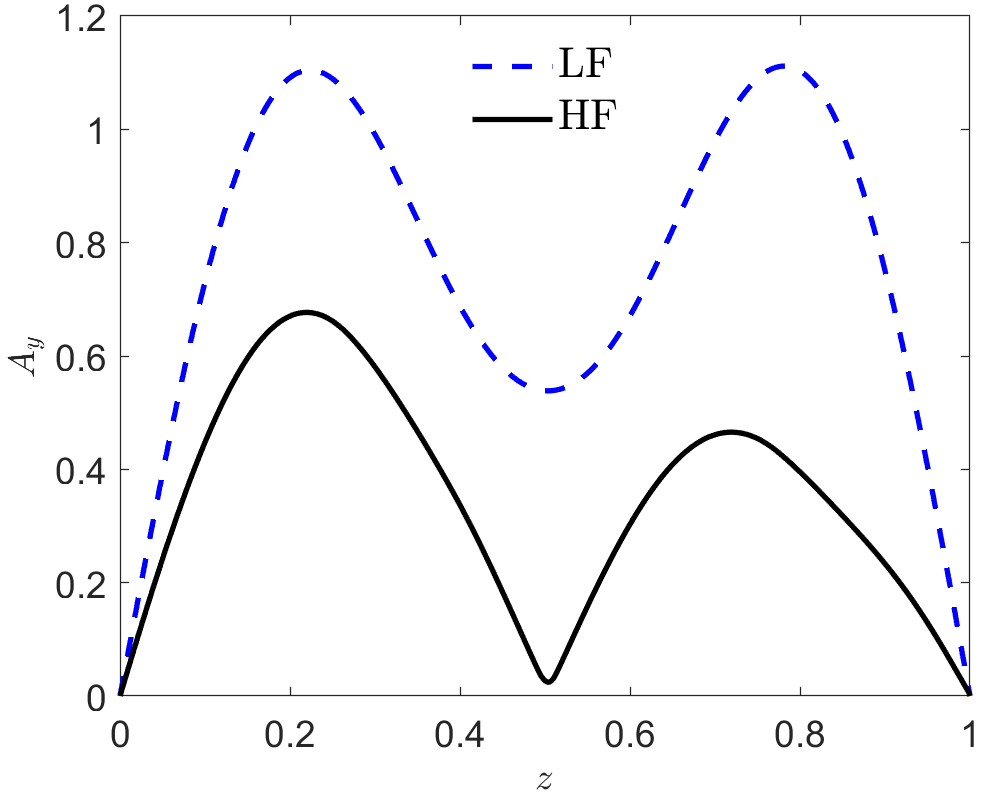}
\includegraphics[width=0.45\textwidth]{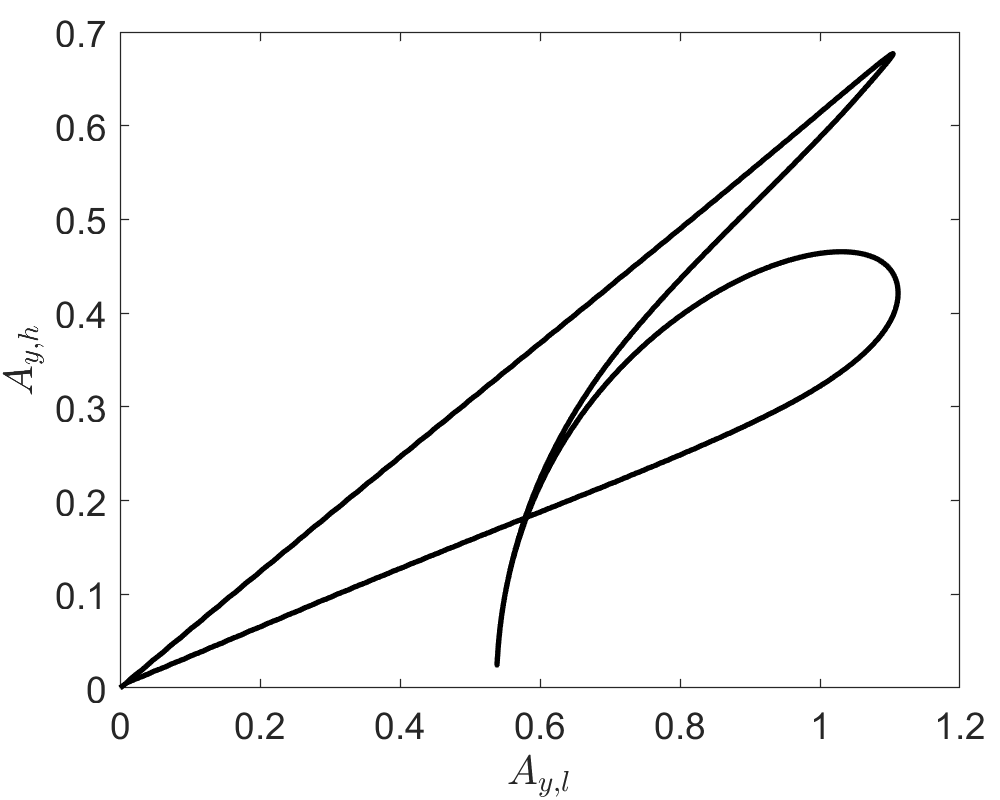}
}
\subfigure[]{\label{fig:exampleb}
\includegraphics[width=0.45\textwidth]{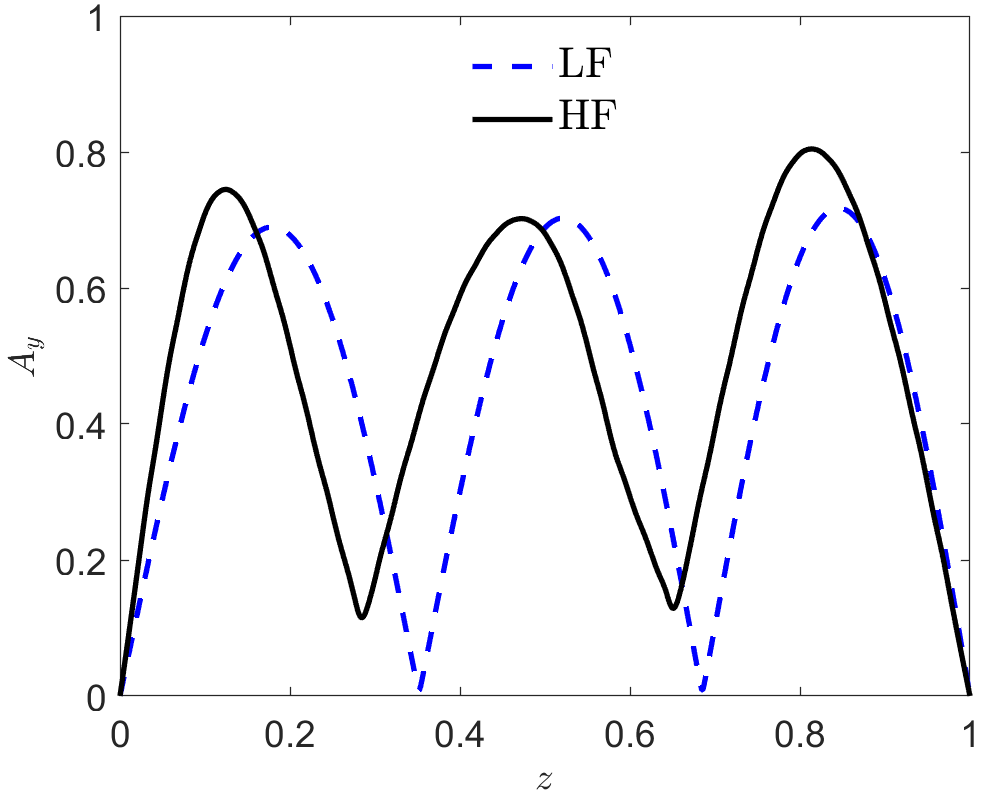}
\includegraphics[width=0.45\textwidth]{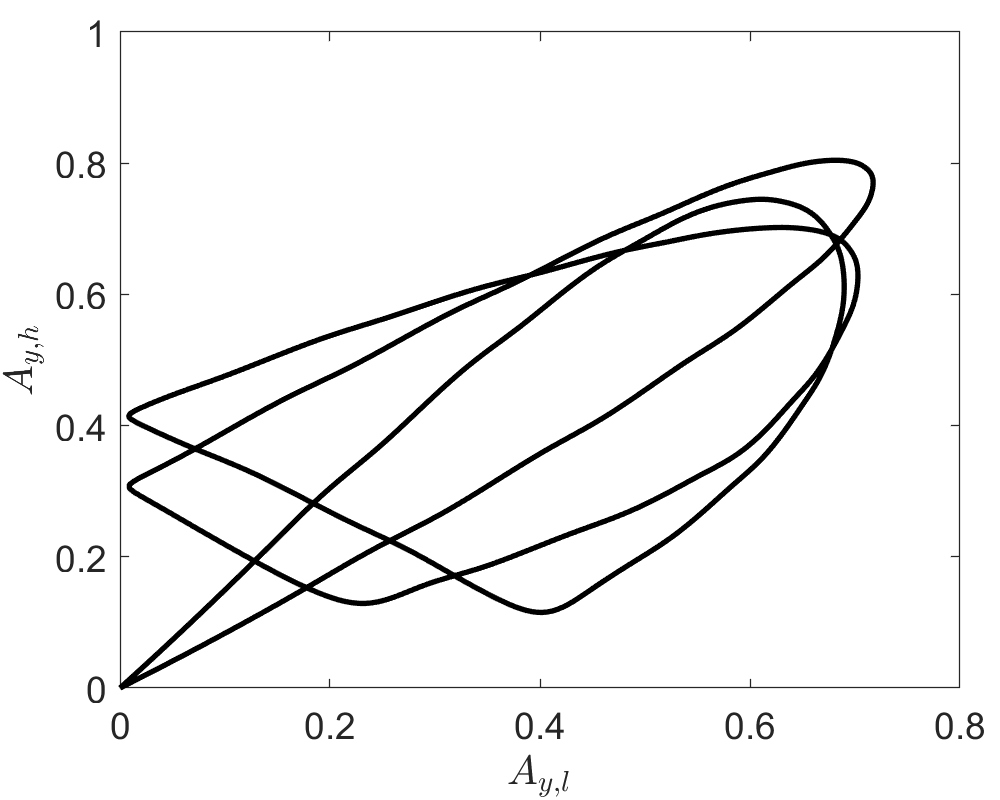}
}
\caption{\label{fig:example}
Correlation between low- and high-fidelity data for two representative cases. 
(a) Uniform flow past a marine riser at $Ur=12.66$. 
Left panel: Low- and high-fidelity data. LF: low-fidelity data that are from VIVA model, HF: high-fidelity data that are  from MIT experiment. 
Right panel: Correlation between the low- and high-fidelity data.
(b) Uniform flow past a marine riser at $Ur=14.54$. 
Left panel: Low- and high-fidelity data. LF: low-fidelity data from the VIVA model, HF: high-fidelity data from the MIT experiment. 
Right panel: Correlation between the low- and high-fidelity data.
$A_y$: displacement, $z$: location, $A_{y,l}$: low-fidelity data (displacements from VIVA), $A_{y,h}$: high-fidelity data (measurements from experiments or field observations). Note that all the displacements ($A_y$/$A_{y,l}/A_{y,h}$) and locations ($z$) in this study are normalized by the diameter  ($d$) and the length of the cylinder ($L$), respectively. 
}
\end{figure}

\begin{figure}
\centering
\subfigure[]{\label{fig:example_fouriera}
\includegraphics[width=0.45\textwidth]{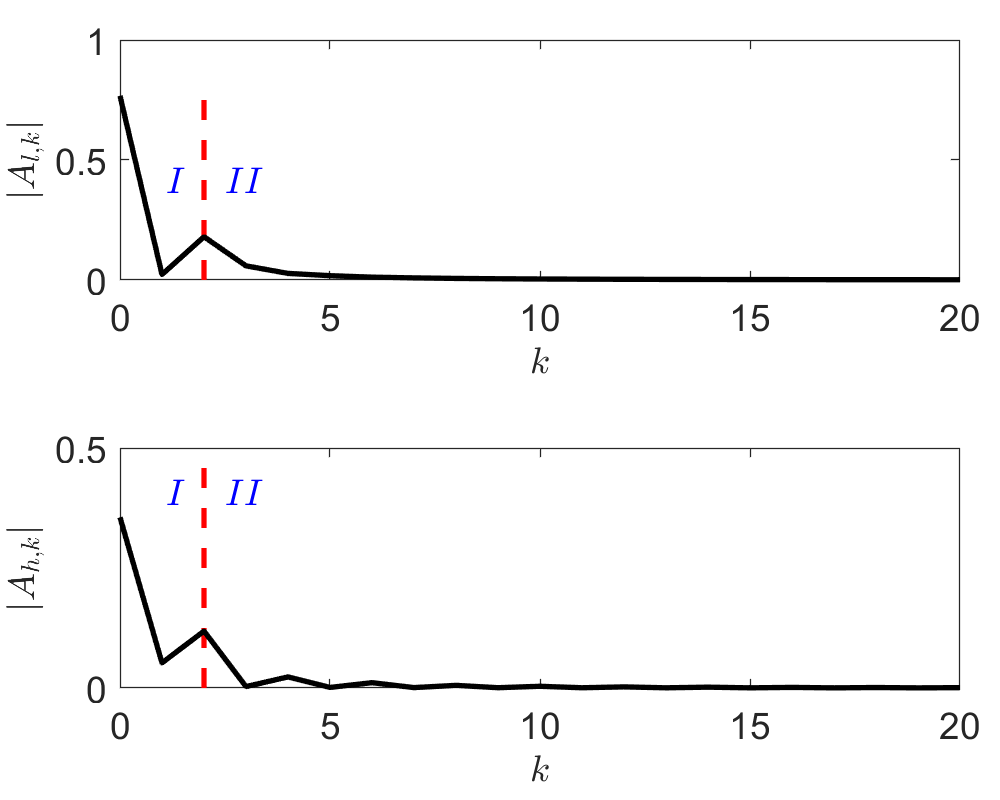}
}
\subfigure[]{\label{fig:example_fourierb}
\includegraphics[width=0.45\textwidth]{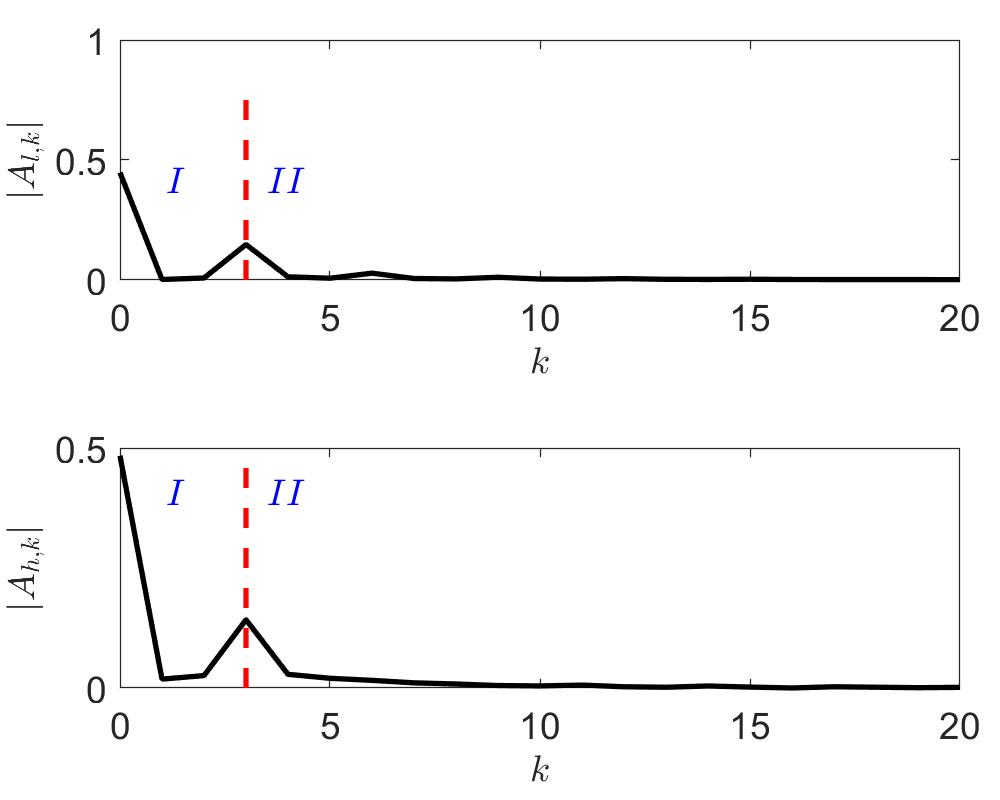}
}
\caption{\label{fig:example_fourier}
Low- and high-fidelity data in the frequency space for two representative cases. 
(a) Uniform flow past a marine riser at $Ur= 12.66$. 
Upper panel: Low-fidelity data in the frequency space. 
Lower panel: Low-fidelity data in the frequency space.
(b) Uniform flow past a marine riser at $Ur= 14.54$. 
Upper panel: Low-fidelity data in the frequency space. 
Lower panel: High-fidelity data in the frequency space.
$k$: wavenumber, $|A_{l,k}|$: module for the $k_{th}$ wave of the low-fidelity data, and $|A_{h,k}|$: module for the $k_{th}$ wave of the high-fidelity data. The red dashed line divides the frequency domain into two sub-domain I and II. The maximal amplitude for waves in I is $O(0.1)$, while it is $O(0.01)$ in II.
}
\end{figure}

Correlations between the low- and high-fidelity models for two representative cases are illustrated as examples in Fig. \ref{fig:example}. Existing multi-fidelity methods perform poorly with such complex correlation functions. 
We observe that there are differences between the low- and high-fidelity data both in the amplitude as well as the spatial phase. Hence, for accurate inference we need to be able to vary the amplitude and phase of the low fidelity data. Therefore, we transform the displacements in the physical space into the modal space using the fast Fourier transformation (FFT) as
\begin{align}
    A_{m,k} = \sum^{N-1}_{n = 0}  A_{y,m,n}  \exp(-i 2 \pi k n /N ), ~ n = 0, ..., N-1,
\end{align}
where $m = l$ or $h$ represents the data at the low- and high-fidelity level, respectively, $A_{m,k} = (\alpha(0), ..., \alpha(k), ...)$ denotes the displacement in the modal space,  $k$ is the wavenumber, $\alpha(k)$ is a complex number denoting the coefficient for the $k_{th}$ wave, and $N$ is the total number of the modes, which is 1,024 in the present work.  As displayed in Fig. \ref{fig:example_fourier}, the low- and high-fidelity data are quite similar in modal space. For example, the module in Fig. \ref{fig:example_fouriera} first decreases and then increases with the increasing $k$ in subdomain I, and then it decreases in subdomain II with increasing $k$. Similar results can also be observed in Fig. \ref{fig:example_fourierb}. Hence, we conduct multi-fidelity modeling in the modal space.  

We now use the case shown in Fig. \ref{fig:example_fouriera} as an example to explain the multi-fidelity model employed in the present study. As observed in the  low-fidelity data of Fig. \ref{fig:example_fouriera}, the modules of the wavenumbers for $k > 6$ are smaller than $0.01$. Therefore, we can keep the first 7 modes and set the remaining modules as equal to zero. Note that the last 6 modes are kept the same as $\alpha(1)-\alpha(6)$ due to the symmetry property of FFT. We then employ the inverse FFT (iFFT) to transform them into the physical space to obtain the low-fidelity function with good accuracy. Due to the similarity of the functional shape of the low- and high-fidelity data in the modal space, we assume that we can accurately predict the  high-fidelity function using the first 7 modes. Predicting the high-fidelity function is now simplified to determining the coefficients for the retained modes in the modal space. 

\begin{algorithm}[H]
\caption{VIV-Multi-fidelity network (VIV-MFnet)}
\label{alg:mf_modal}
\begin{algorithmic}
\Require Low- (VIVA simulations) and high-fidelity (Experiment/Field observations) data.

\State $\bullet$ Perform FFT for low-fidelity data  to obtain $A_{l,k}$.\;
\State $\bullet$ Obtain $k_c$ and $k_t$ based on the low-fidelity data in modal space.\;
\State $\bullet$ Set $\alpha(k)$ as zero for $k > k_t$.\;
\State $\bullet$ Optimize $\mathcal{L}(\bm{\alpha})$ in Eq. \eqref{eq:loss_2} using the Adam optimizer to obtain $\alpha(k)$ for $k \le k_t$.\;
\State $\bullet$ Conduct HMC to get samples for $\alpha(k)$ from $P(\bm{\alpha}|D)$ in Eq. \eqref{eq:bayes}.\;
\State $\bullet$ Get predictions $\hat{A}_{y,h}$ from Step 4 or 5.
\end{algorithmic}
\end{algorithm}

\subsection{Maximum a posteriori probability (MAP) estimate}
\label{sec:map}
To obtain the coefficients in the modal space, we can minimize the following loss function:
\begin{align}\label{eq:loss}
    \mathcal{L}(\bm{\alpha}) = |A_{y,h}^* - \hat{A}_{y,h}|^2 + \lambda |\bm{\alpha}|^2, 
\end{align}
where $A_{y,h}^*$ represents the high-fidelity measurements, $\hat{A}_{y,h}$ is the high-fidelity prediction obtained from the iFFT, and the second term is the $\mathcal{L}_2$ regularization used to prevent overfitting. In addition, $\bm{\alpha}$ denotes the coefficients in the modal space, and $\lambda$ is the weight for $\bm{\alpha}$.


The entire frequency space is divided into two subdomains, I and II, as shown in Fig. \ref{fig:example_fourier}. We observe that the maximum modules for the low-fidelity data in subdomains I and II are of order $O(0.1)$ and $O(0.01)$, respectively.  We assume that this also holds for the high-fidelity data, and then utilize different weights for the coefficients in these two subdomains. Specifically, we can rewrite Eq. \eqref{eq:loss} as follows:
\begin{align}\label{eq:loss_2}
    \mathcal{L}(\bm{\alpha}) = |A_{y,h}^* - \hat{A}_{y,h}|^2 + \lambda_{I} |\bm{\alpha}_{I}|^2 + \lambda_{II} |\bm{\alpha}_{II}|^2, 
\end{align}
where $\lambda_{I} (\bm{\alpha}_{I})$ and $\lambda_{II} (\bm{\alpha}_{II})$ are the weights in subdomain I and II, respectively. In addition, the first term in Eq. \eqref{eq:loss_2} is generally as small as O($10^{-4}$) based on our experience. Therefore, we can set $\lambda_{I} \thicksim O(0.01)$ and $\lambda_{II} \thicksim O(0.1)$ or $\lambda_{II} \thicksim O(1)$ to guarantee that the three terms in Eq. \eqref{eq:loss_2} are of the similar order, i.e., O($10^{-4}$). Note that no regulation is employed for $k = 0$. 

The choice of $\lambda_I$ and $\lambda_{II}$ is crucial to ensure accurate inference.  Generally, we can set $\lambda_{I} = 0.01$ and $\lambda_{II} = \lambda_{I} |\alpha_{I,max}|/|\alpha_{II,max}|$, where $\alpha_{I,max}$ and $\alpha_{II,max}$ are the maxima in subdomain I and II, respectively.  In particular, we can tune $\lambda_{II}$ based on $\lambda_{I}$ and the ratio between the maximum modules in subdomain I and II. In addition, the critical wavenumber between the two subdomains is denoted as $k_c$, and the truncation wavenumber is denoted as $k_t$.  We note that the values of $\alpha_{I,max}$, $\alpha_{II,max}$, $k_c$, and $k_t$ are determined from the low-fidelity data.
Specifically, we employ $\lambda_I = 0.01$ and $\lambda_{II} = 0.25$, which works well for all cases studied in this work.

\subsection{Bayesian inference (BI) for uncertainty quantification}
\label{sec:bayes}
In the Bayesian framework, each coefficient is a  distribution rather than the point estimate in Sec. \ref{sec:map}. Specifically,  by denoting the observed data as $\mathcal{D}$,  the posterior distribution of $\bm{\alpha}$ can be expressed as follows, based on Bayes' rule:
\begin{align}\label{eq:bayes}
    P(\bm{\alpha}|\mathcal{D}) = \frac{P(\bm{\alpha}) P(\mathcal{D}|\bm{\alpha})}{P(\mathcal{D})},  
\end{align}
where $P({\bm{\alpha}})$ is the prior distribution of the unknown coefficients,  and $P(\mathcal{D}|\bm{\alpha})$ is the  likelihood.  $P(\bm{\alpha}|\mathcal{D})$ is basically intractable since $P(\mathcal{D})$ is unknown. However, we can  use the Hamiltonian Monte Carlo (HMC) \cite{neal2011mcmc,betancourt2017conceptual} to sample from $P(\bm{\alpha}|\mathcal{D})$.  Similar as the regularization in Sec. \ref{sec:map}, we divide the prior $P({\bm{\alpha}})$ into two parts and set different priors for each subdomain, to prevent overfitting. In particular, we employ $P({\bm{\alpha}_I}) \thicksim \mathcal{N}(0, 0.1^2)$ and $P({\bm{\alpha}_{II}}) \thicksim \mathcal{N}(0, 0.01^2)$ in this study. 
We would like to point out that MAP is computationally very efficient but does not predict uncertainty, while BI is relatively more expensive but is capable of handling noisy data as well as quantifying uncertainty in predictions. 

\section{Results and Discussion}
\label{sec:results}
In this section, we first apply the approach described in Sec. \ref{sec:map}, i.e., MAP, to predict riser displacements in uniform and sheared flows. In particular, we conduct a comparison between the present method and the state-of-art multi-fidelity approach, MF-DNNs \cite{meng2020composite}, which is capable of capturing nonlinear cross-correlation between the low-and high-fidelity data. We then employ the method described in Sec. \ref{sec:bayes}, i.e., BI, to quantify uncertainty in our predictions. In MAP, the Adam optimizer with an initial learning rate is set as $10^{-3}$, and the number of  training steps is 10,000. In HMC, we set the number of burn-in steps as 1,000 with the leapfrog step 20; also, the initial time step is set as $0.1$. Finally, 1,000 samples are employed to compute the mean and standard deviation after the burn-in steps.

\subsection{Results from MAP}
\label{sec:map_result}

\subsubsection{Lab experiments for uniform and shear flows}
\begin{figure}
\centering
\subfigure[]{\label{fig:mf_2a}
\includegraphics[width=0.3\textwidth]{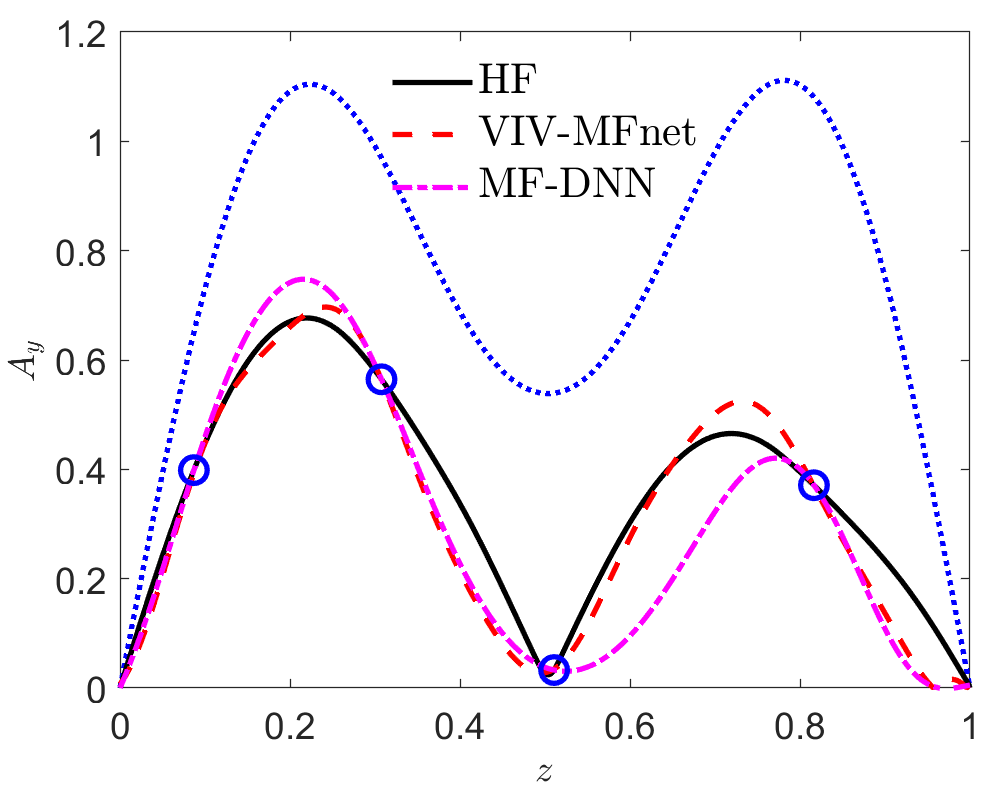}}
\subfigure[]{\label{fig:mf_2b}
\includegraphics[width=0.3\textwidth]{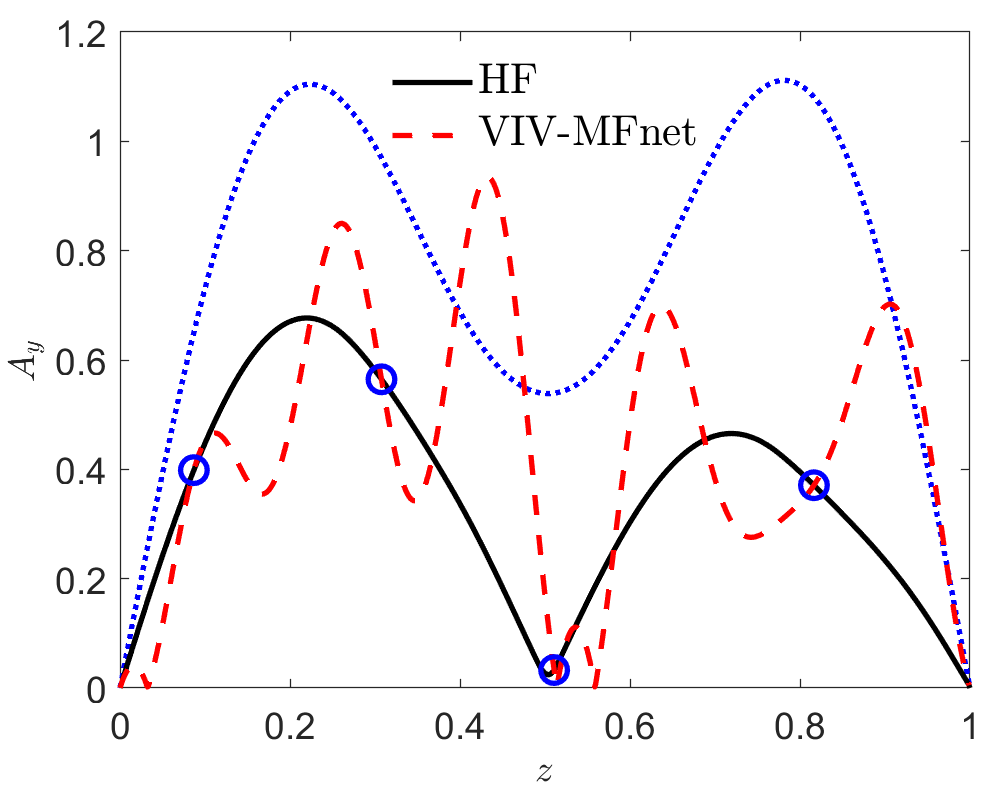}}
\subfigure[]{\label{fig:mf_2c}
\includegraphics[width=0.3\textwidth]{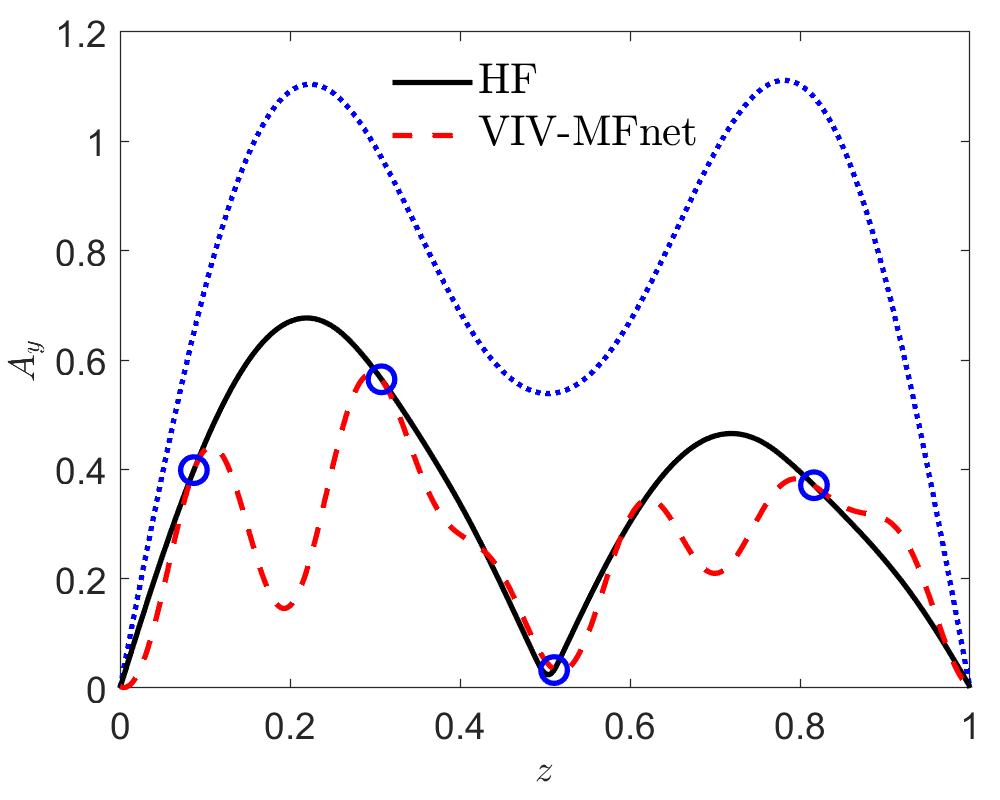}}
\caption{\label{fig:mf_2}
Importance of regularization: Displacements for uniform flow past a marine riser at $Ur=12.66$.
(a) Multi-fidelity predictions with $\mathcal{L}_2$ regularization $\lambda_I = 0.01$, and $\lambda_{II} = 0.25$. 
(b) Multi-fidelity predictions with no $\mathcal{L}_2$ regularization, i.e.,  $\lambda_I = \lambda_{II} = 0$. 
(c) Multi-fidelity predictions with $\mathcal{L}_2$ regularization $\lambda_I = \lambda_{II} = 0.01$. Blue dot: low-fidelity training data, Blue circle: high-fidelity training data. HF: MIT experiment, VIV-MFnet: multi-fidelity modeling in Sec. \ref{sec:map}, MF-DNN: multi-fidelity modeling using \cite{meng2020composite}.
The first DNN in MF-DNNs has 2 hidden layers with 40 neurons per layer, and the second DNN in MF-DNNs has 2 hidden layers with 20 neurons per layer. The hyperbolic tangent function is employed as the activation function in both DNNs. The Adam optimizer is used for optimization. More details for MF-DNNs can be found in \cite{meng2020composite}. 
}
\end{figure}

We fist test the case with uniform flow at $Ur = 12.66$. The critical and truncated wavenumbers for this case are $k_c = 2$ and $k_t = 6$, respectively, according to Fig. \ref{fig:example_fouriera}. In our multi-fidelity modeling, we assume that we have 4 high-fidelity measurements, which are randomly distributed in $z \in [0, 1]$ as displayed in Fig. \ref{fig:mf_2}. The predicted high-fidelity profile with $\lambda_I = 0.01$ and $\lambda_{II} = 0.25$ is illustrated in Fig. \ref{fig:mf_2a}, which agrees well with the experimental results.  In particular,  predicting the maximum displacement is significant in the design of riser, because it determines the stress level: the maximum displacements for the multi-fidelity modeling in Fig. \ref{fig:mf_2a} and the experiment are 0.6968 and 0.6769, respectively. 
In addition, we use the recently developed multi-fidelity deep neural networks because they are highly expressive in capturing nonlinear correlation between the low- and high-fidelity data \cite{meng2020composite}. Here we also present the results using the MF-DNNs in Fig. \ref{fig:mf_2a} for comparison. As shown, the present method outperforms the MF-DNNs, especially for predictions in the interval $z \in [0.5, 0.7]$.

Next, we study the effect of $\mathcal{L}_2$ regularization on the accuracy of the inference. Specifically, two additional scenarios are considered, viz., no $\mathcal{L}_2$ regularization is employed in either subdomain, and the same weight ($\bm{\lambda} > 0$) is used in both subdomains.   We observe that: (1) it is easy to get overfitting if no $\mathcal{L}_2$ regularization is used (Fig. \ref{fig:mf_2b}), and (2) the usage of different weights for waves in different subdomains can enhance the predicted accuracy when comparing the results in Fig. \ref{fig:mf_2c} to those in Fig. \ref{fig:mf_2a}.

\begin{figure}
\centering
\subfigure[]{\label{fig:mf_3a}
\includegraphics[width=0.3\textwidth]{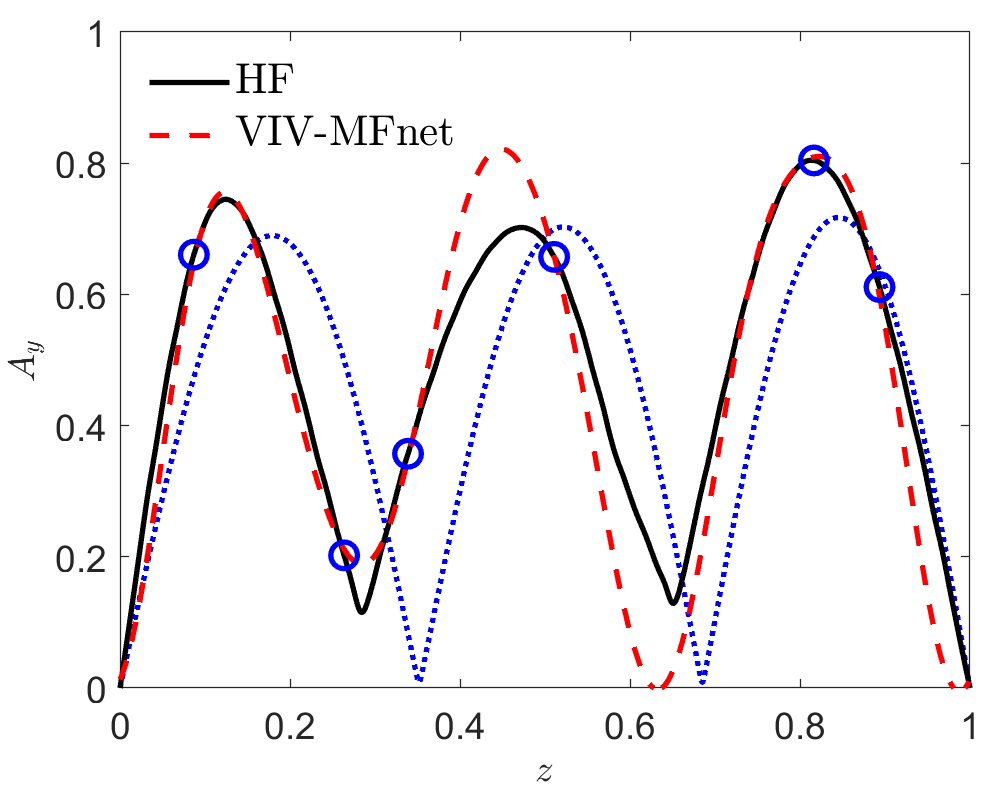}}
\subfigure[]{\label{fig:mf_3b}
\includegraphics[width=0.3\textwidth]{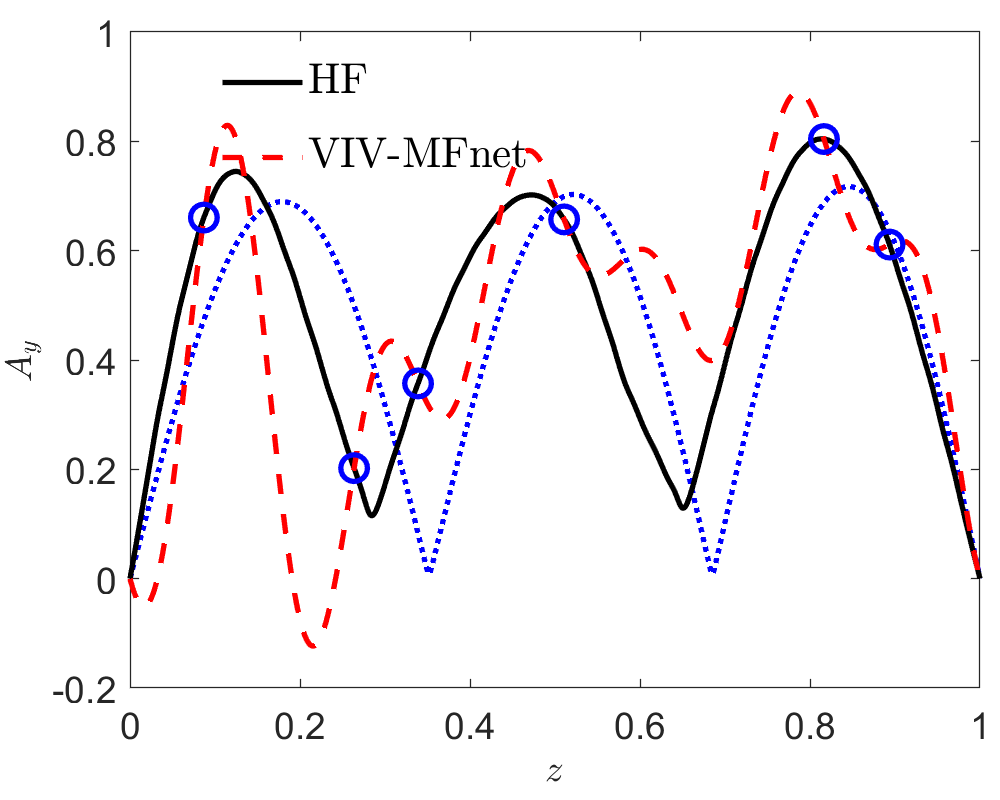}}
\caption{
Multi-fidelity predictions of displacements for uniform flow past a marine riser at $Ur=14.54$.
(a) With $\mathcal{L}_2$ regularization $\lambda_I = 0.01$, and $\lambda_{II} = 0.25$. 
(b)  With no $\mathcal{L}_2$ regularization, i.e.,  $\lambda_I = \lambda_{II} = 0$. 
 HF: MIT experiment.
 Blue dot: low-fidelity training data, Blue circle: high-fidelity training data.
}
\label{fig:mf_3}
\end{figure}

We now consider the case in Fig. \ref{fig:exampleb}, in which the critical and truncated wavenumbers are set as $k_c = 3$ and $k_t = 6$, respectively, based on Fig. \ref{fig:example_fourierb}. Similarly, we perform two tests with different regularizations, i.e., (1) $\lambda_I = 0.01$, $\lambda_{II} = 0.25$, and (2) $\lambda_I = \lambda_{II} = 0$. We assume that we have 6 random samples for training data here. As shown in Fig. \ref{fig:mf_3}, the predictions from multi-fidelity modeling with $\mathcal{L}_2$ regularization are in good agreement with the experimental results, while overfitting is observed in the results without $\mathcal{L}_2$ regularization.  

\begin{figure}
\centering
\subfigure[]{\label{fig:mf_4a}
\includegraphics[width=0.45\textwidth]{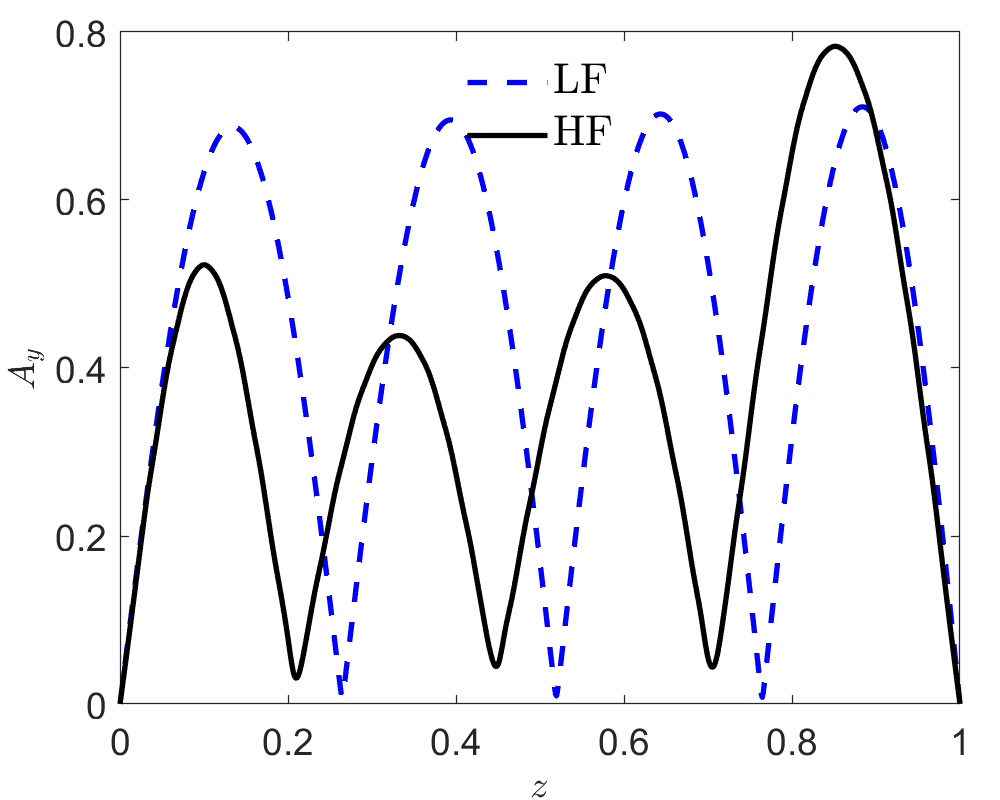}
\includegraphics[width=0.45\textwidth]{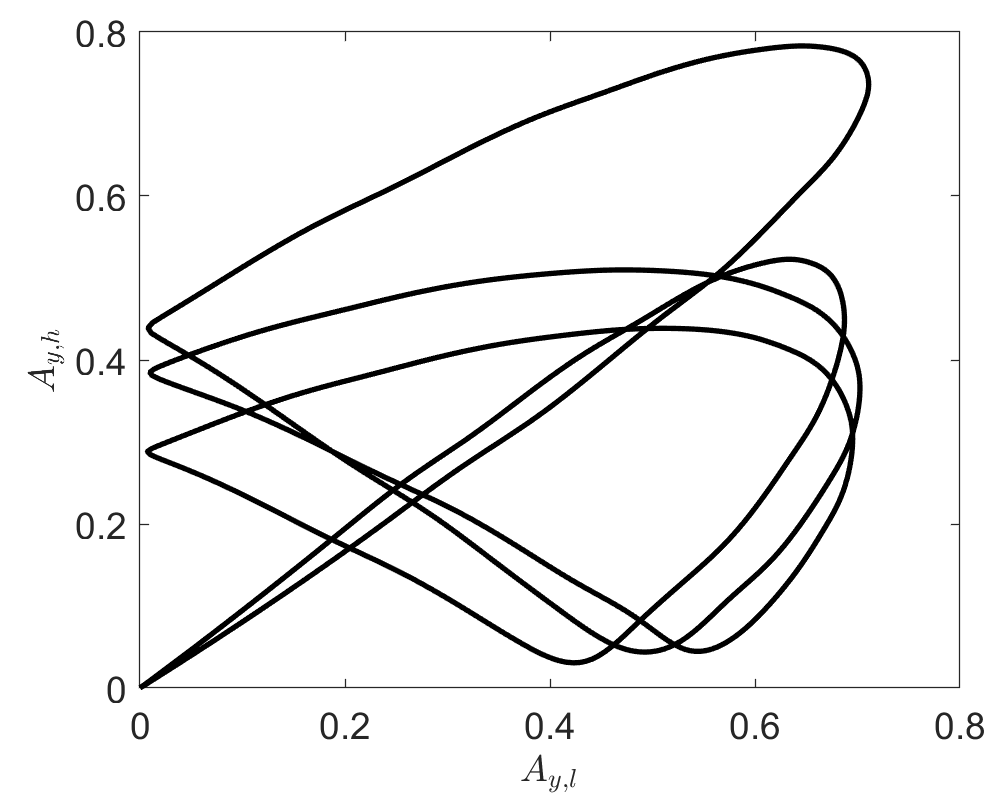}
}
\subfigure[]{\label{fig:mf_4b}
\includegraphics[width=0.45\textwidth]{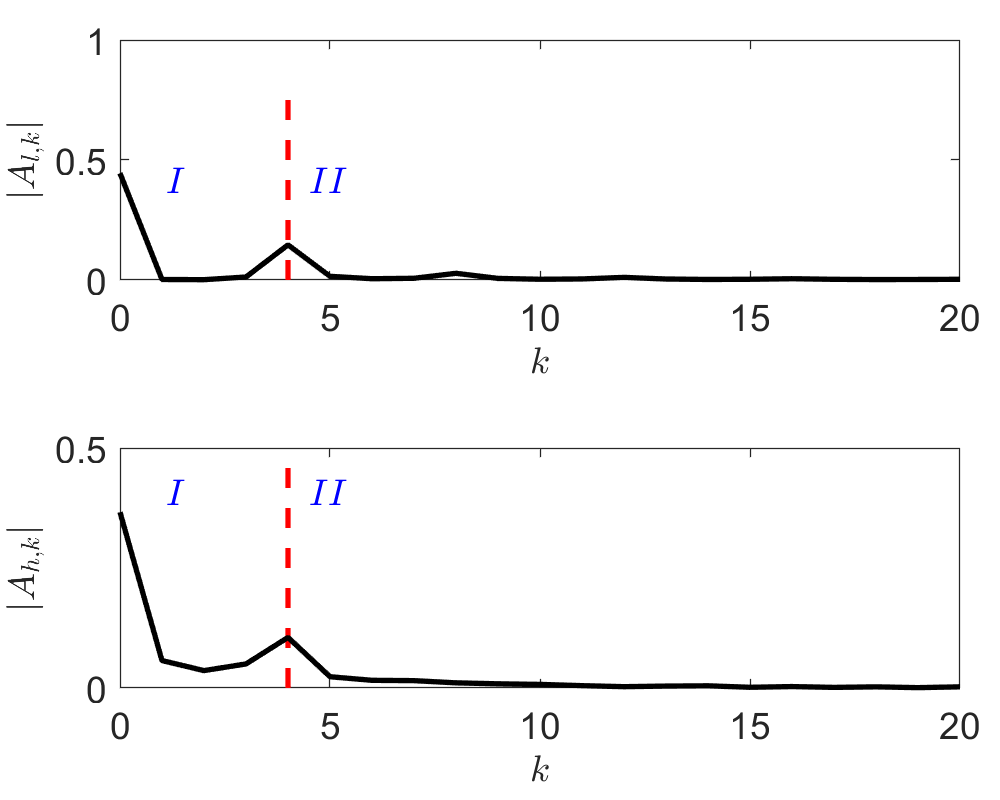}
\includegraphics[width=0.45\textwidth]{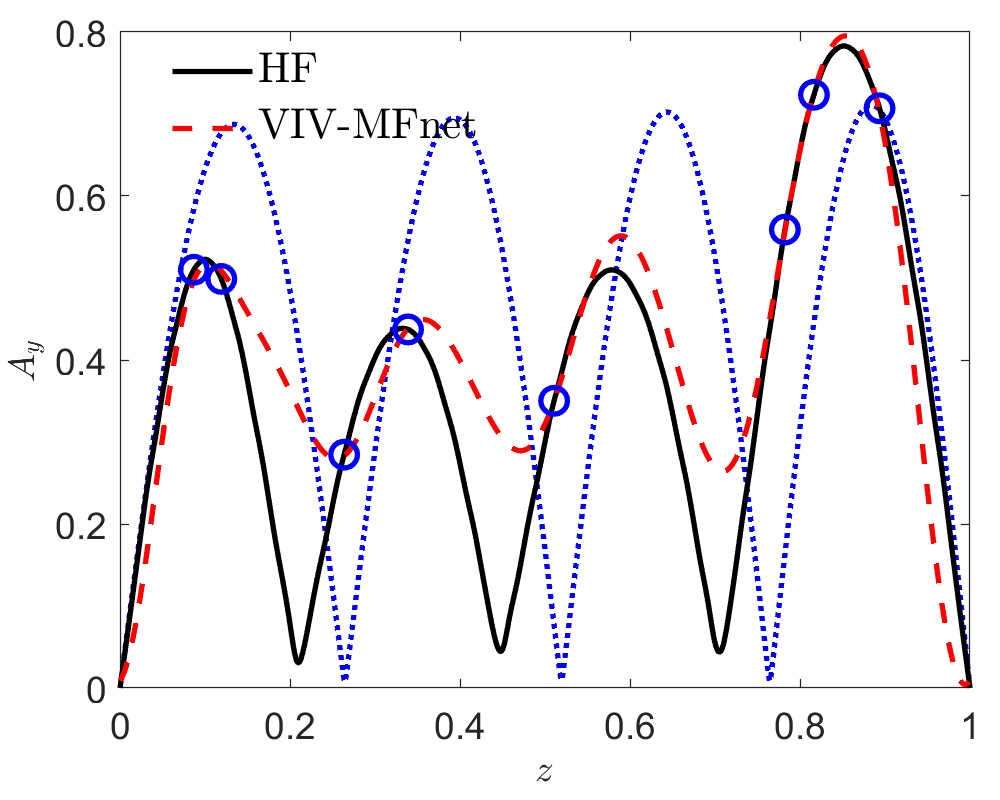}
}
\caption{\label{fig:mf_4}
Multi-fidelity predictions of displacements for uniform flow past a marine riser at $Ur=21.64$.
(a)  Left panel: Low- and high-fidelity data. LF: low-fidelity data from the VIVA model, HF: high-fidelity data from the MIT experiment. 
Right panel: Correlation between the low- and high-fidelity data.
(b) Left panel: Low- and high-fidelity data in the frequency space. First row: low-fidelity data, Second row: high-fidelity data. The critical and truncated wavenumbers are $k_c = 4$ and $k_t = 9$, respectively. 
Right panel: With $\lambda_I = 0.01$, and $\lambda_{II} = 0.25$. 
Blue dot: low-fidelity training data, Blue circle: high-fidelity training data.
}
\end{figure}

\begin{figure}
\centering
\subfigure[]{\label{fig:mf_5a}
\includegraphics[width=0.45\textwidth]{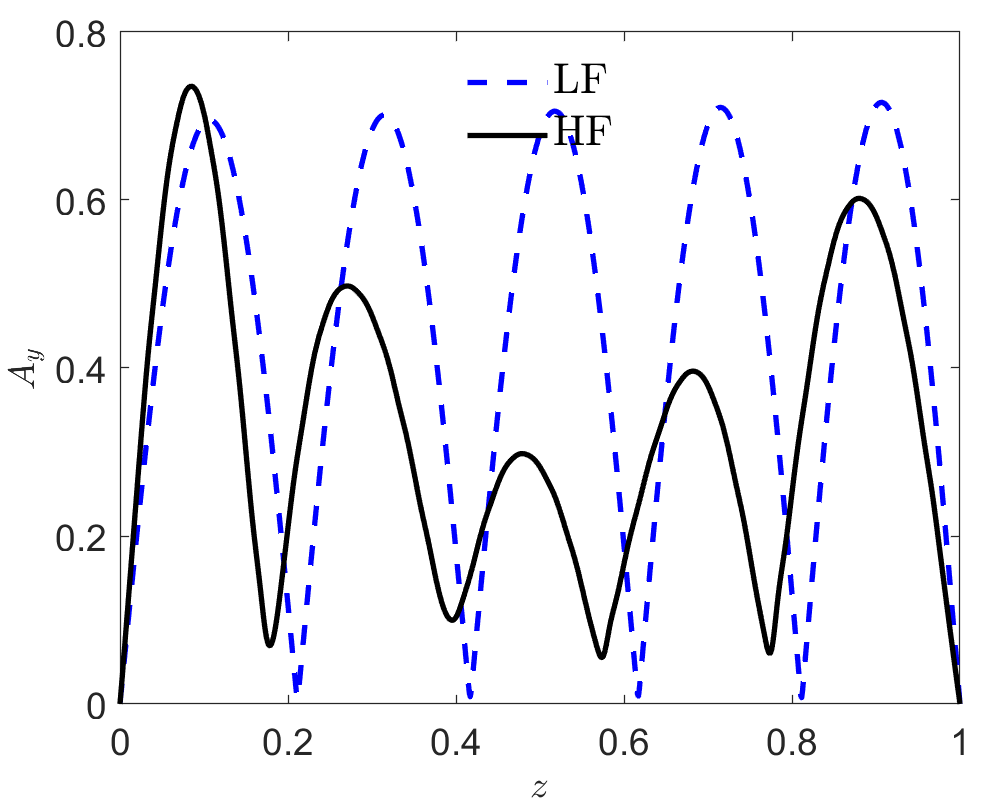}
\includegraphics[width=0.45\textwidth]{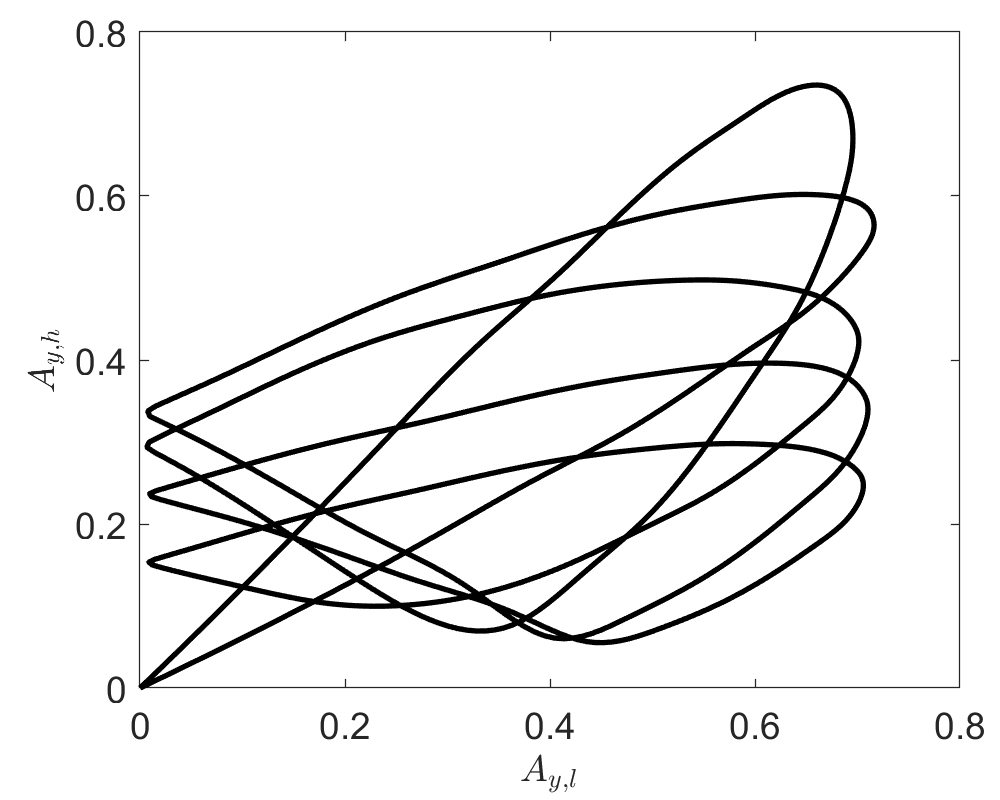}
}
\subfigure[]{\label{fig:mf_5b}
\includegraphics[width=0.45\textwidth]{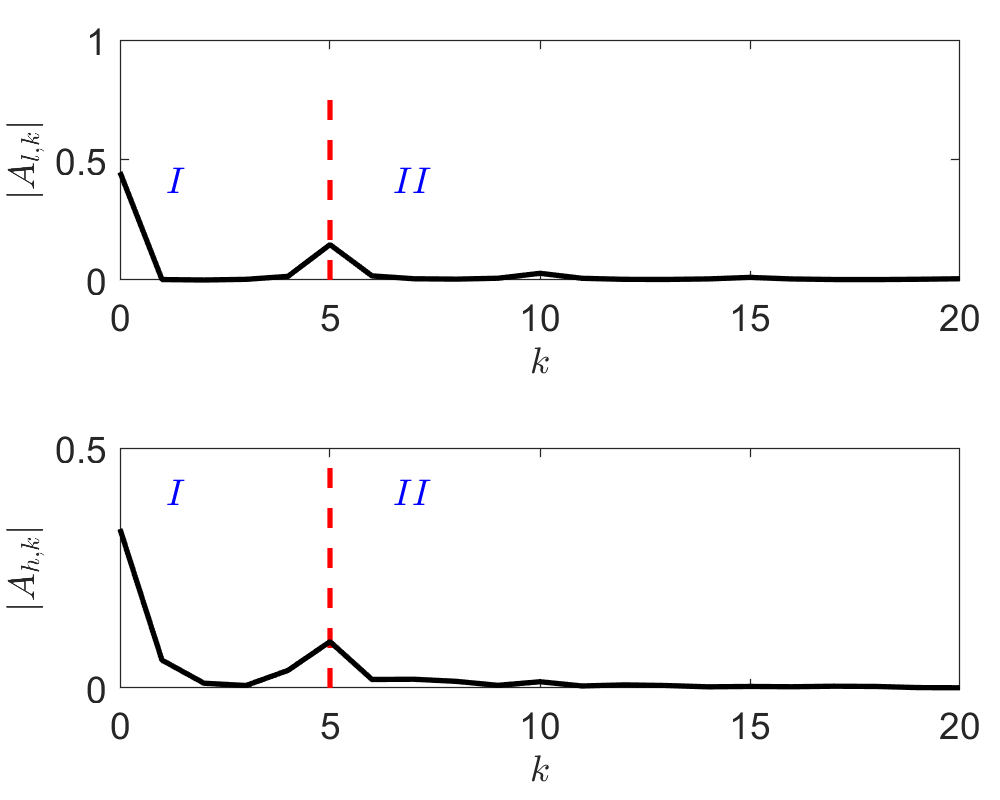}
\includegraphics[width=0.45\textwidth]{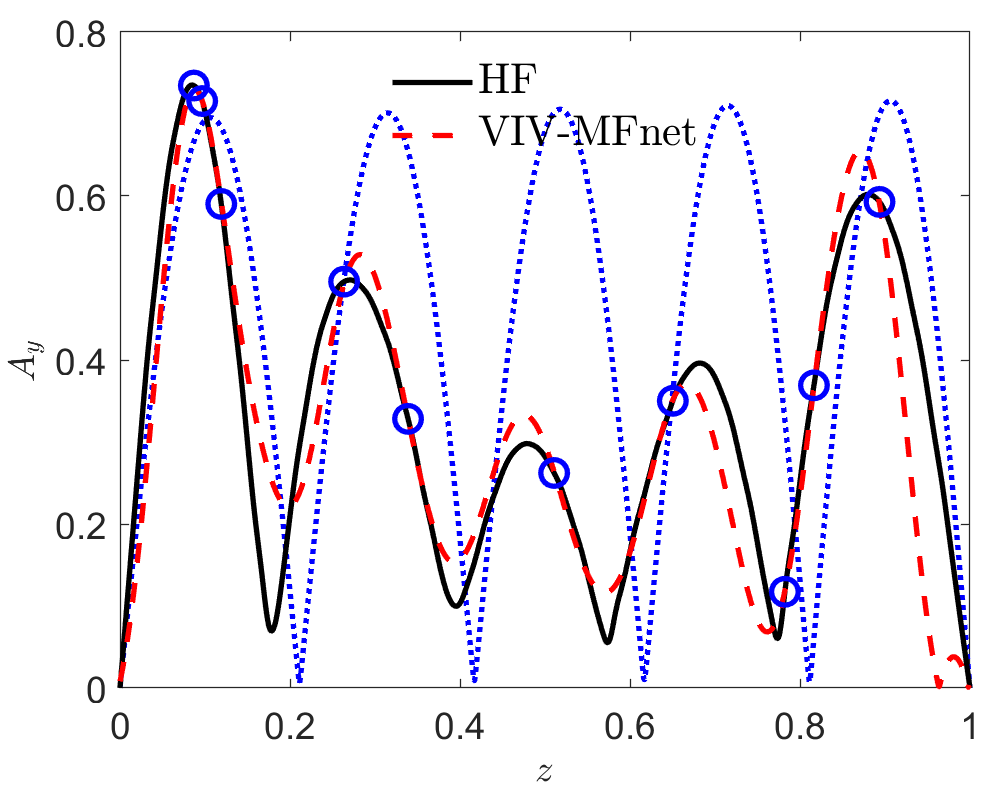}
}
\caption{\label{fig:mf_5}
Multi-fidelity predictions of displacements for uniform flow past a marine riser at $Ur=26.26$.
(a)  Left panel: Low- and high-fidelity data. LF: low-fidelity data from the VIVA model, HF: high-fidelity data from the MIT experiment. 
Right panel: Correlation between the low- and high-fidelity data.
(b) Left panel: Low- and high-fidelity data in the frequency space. First row: low-fidelity data, Second row: high-fidelity data. The critical and truncated wavenumbers are $k_c = 5$ and $k_t = 12$, respectively. 
Right panel: With $\lambda_I = 0.01$, and $\lambda_{II} = 0.25$. 
Blue dot: low-fidelity training data, Blue circle: high-fidelity training data.
}
\end{figure}

Next, we test the performance of the proposed method for predicting displacements of the riser within uniform flow at different velocities, i.e., $Ur = 21.64$ and $26.26$; 8 and 10 random samples are employed as the training data for these two cases, respectively. As shown in Figs. \ref{fig:mf_4}-\ref{fig:mf_5}, the multi-fidelity predictions agree well with the experimental results.

\begin{figure}
\centering
\subfigure[]{\label{fig:mf_3_4a}
\includegraphics[width=0.3\textwidth]{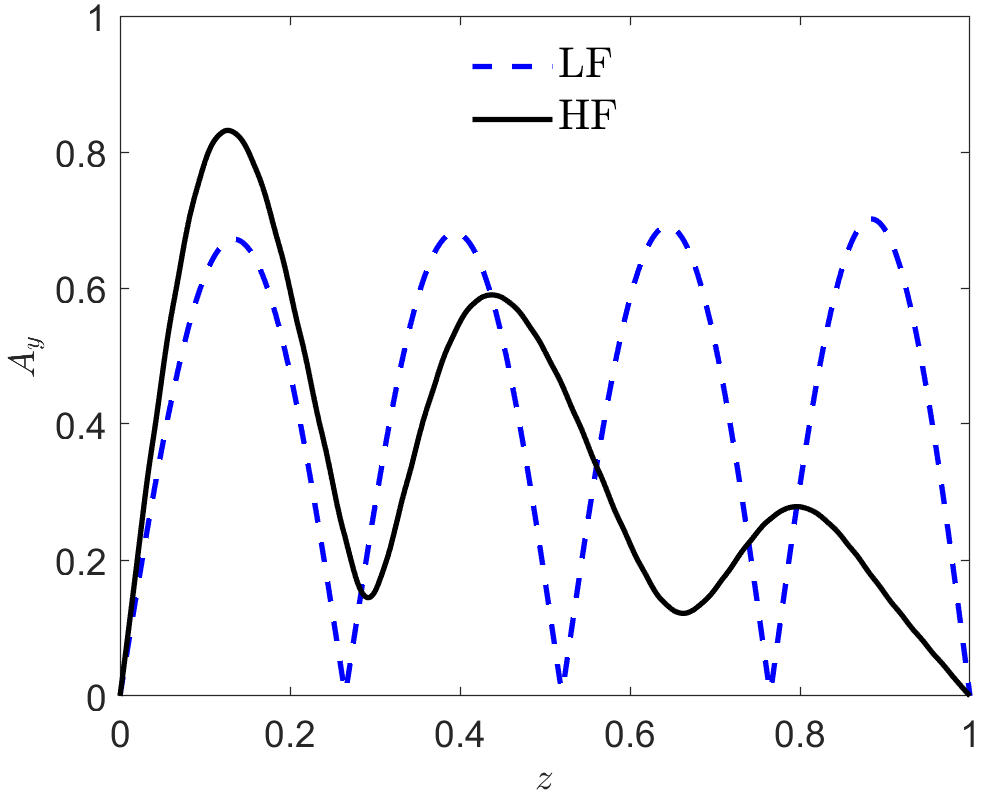}
\includegraphics[width=0.3\textwidth]{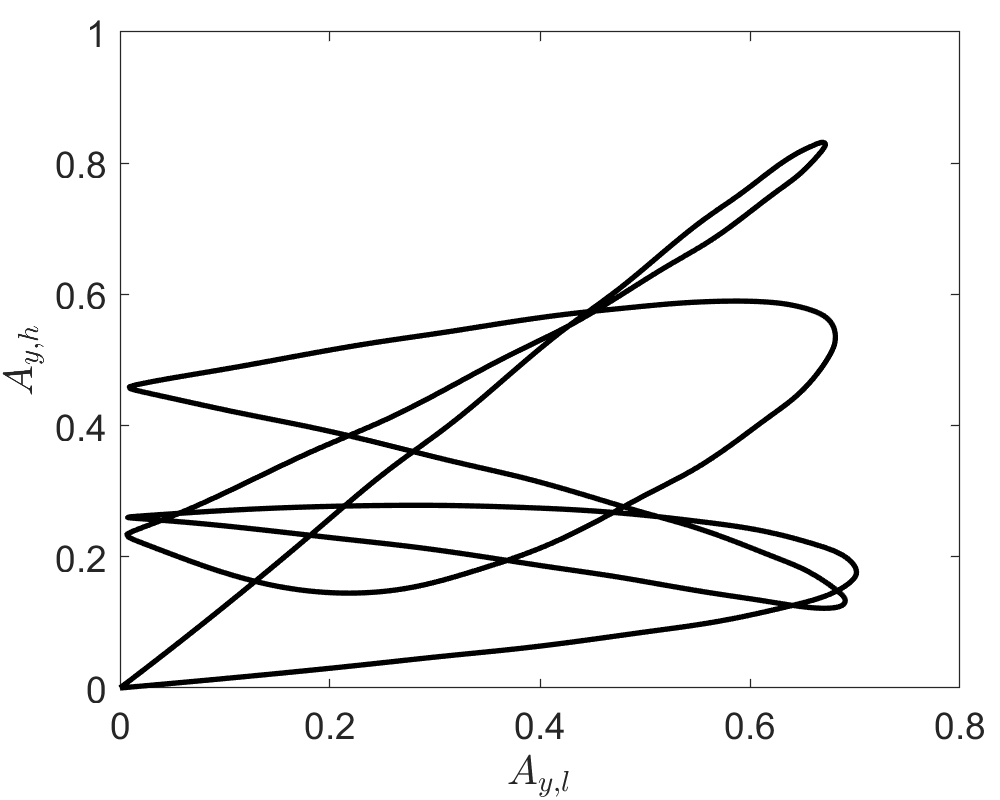}
\includegraphics[width=0.3\textwidth]{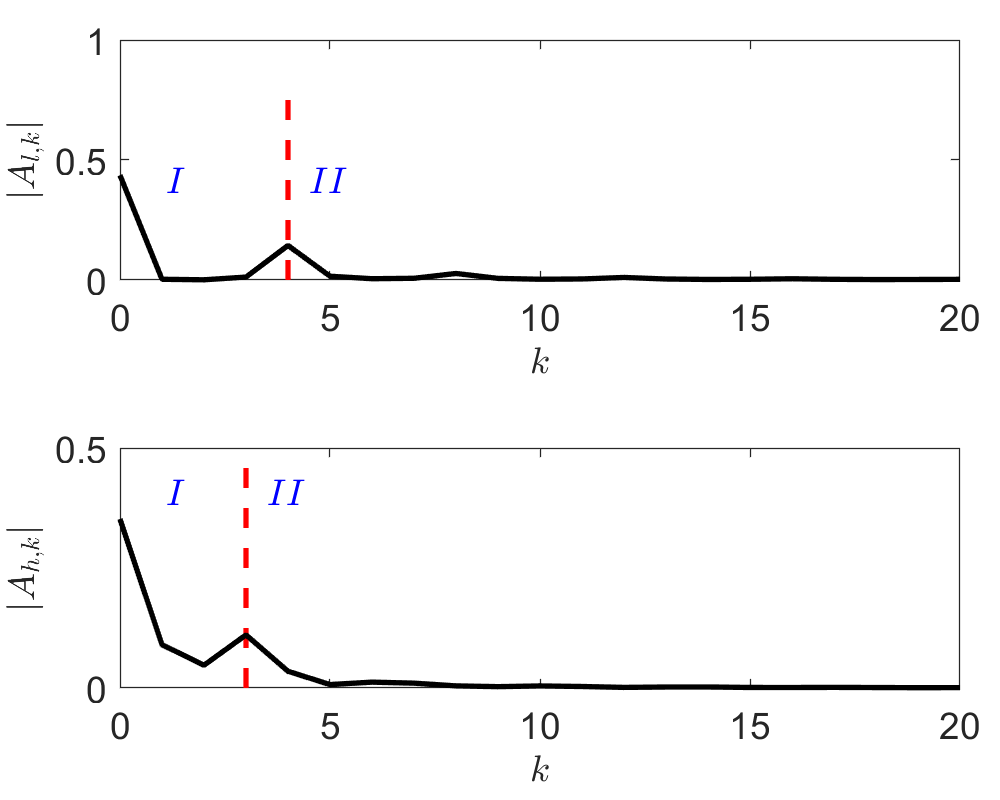}
}
\subfigure[]{\label{fig:mf_3_4b}
\includegraphics[width=0.3\textwidth]{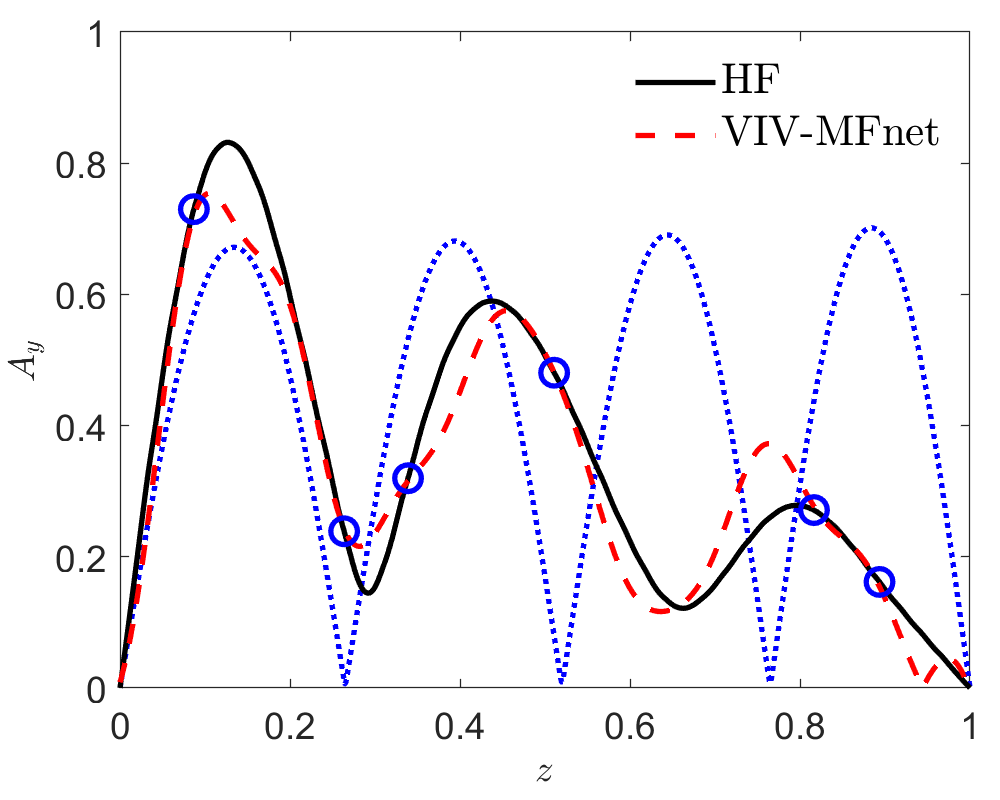}
\includegraphics[width=0.3\textwidth]{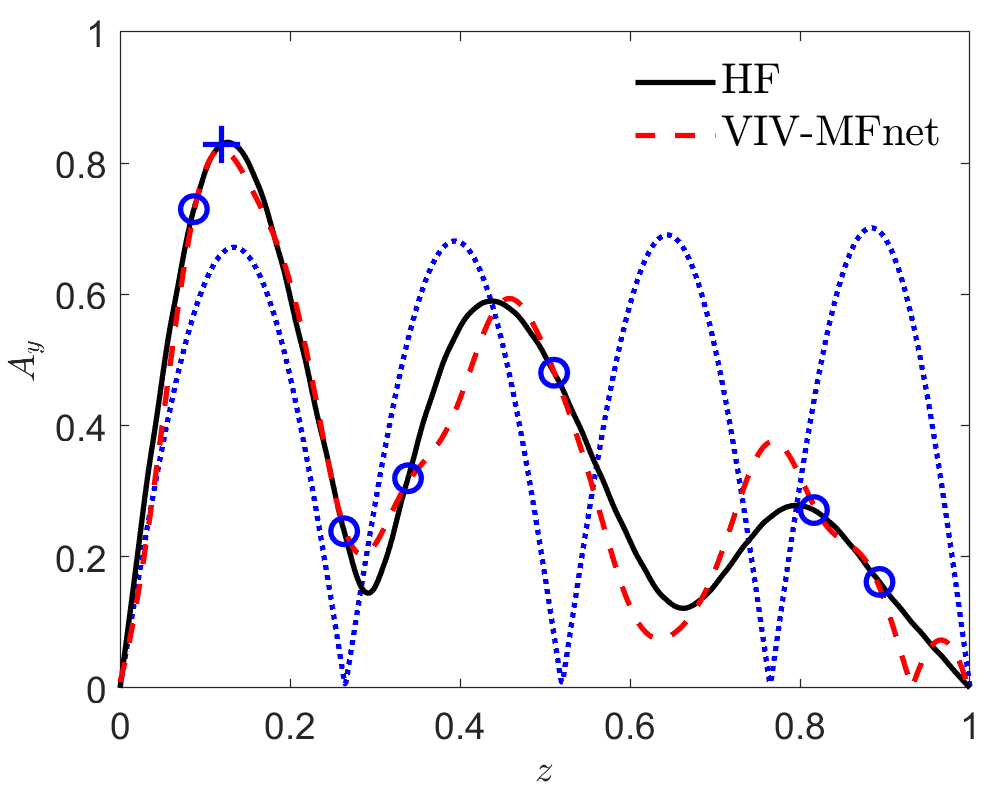}
}
\caption{\label{fig:mf_3_4}
Multi-fidelity predictions of displacements for uniform flow past a marine riser at $Ur=17.23$.
(a)  Left panel: Low- and high-fidelity data. LF: low-fidelity data that are from VIVA model, HF: high-fidelity data that are from MIT experiment. 
Middle panel: Correlation between the low- and high-fidelity data.
Right panel: Low- and high-fidelity data in the frequency space. First row: low-fidelity data, Second row: high-fidelity data. Note that the critical wavenumber is different in the low- (i.e., $k_c = 4$) and high-fidelity (i.e., $k_c = 3$) data.
(b)  Left panel: With 6 random training data. 
Right panel: With 7 random training data.  In both cases, $\lambda_I = 0.01$, and $\lambda_{II} = 0.25$. 
Blue dot: low-fidelity training data, Blue cross: high-fidelity training data, Blue cross: added high-fidelity training data.
}
\end{figure}

\begin{figure}
\centering
\subfigure[]{\label{fig:mf_7a}
\includegraphics[width=0.45\textwidth]{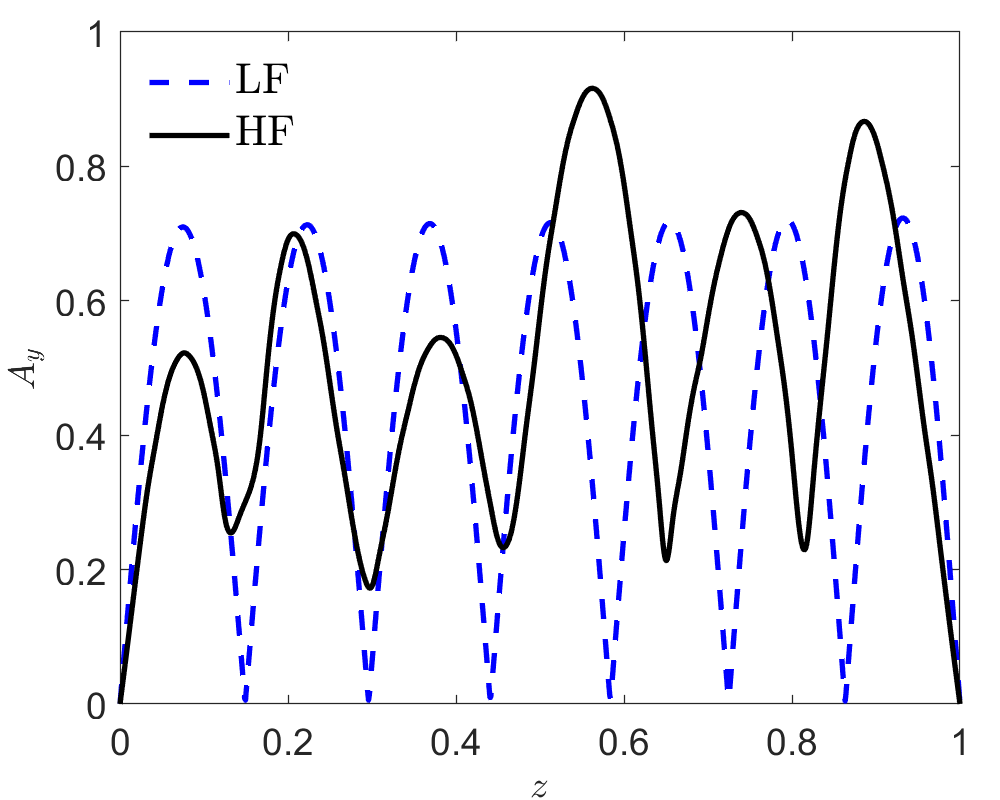}
\includegraphics[width=0.45\textwidth]{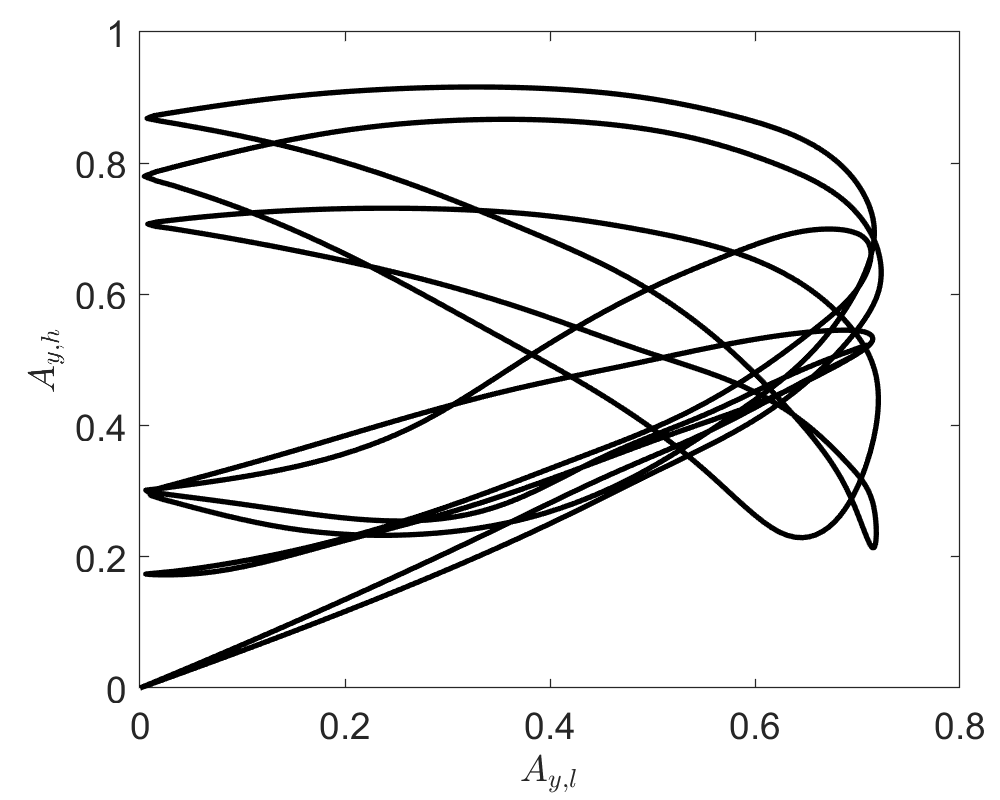}
}
\subfigure[]{\label{fig:mf_7b}
\includegraphics[width=0.45\textwidth]{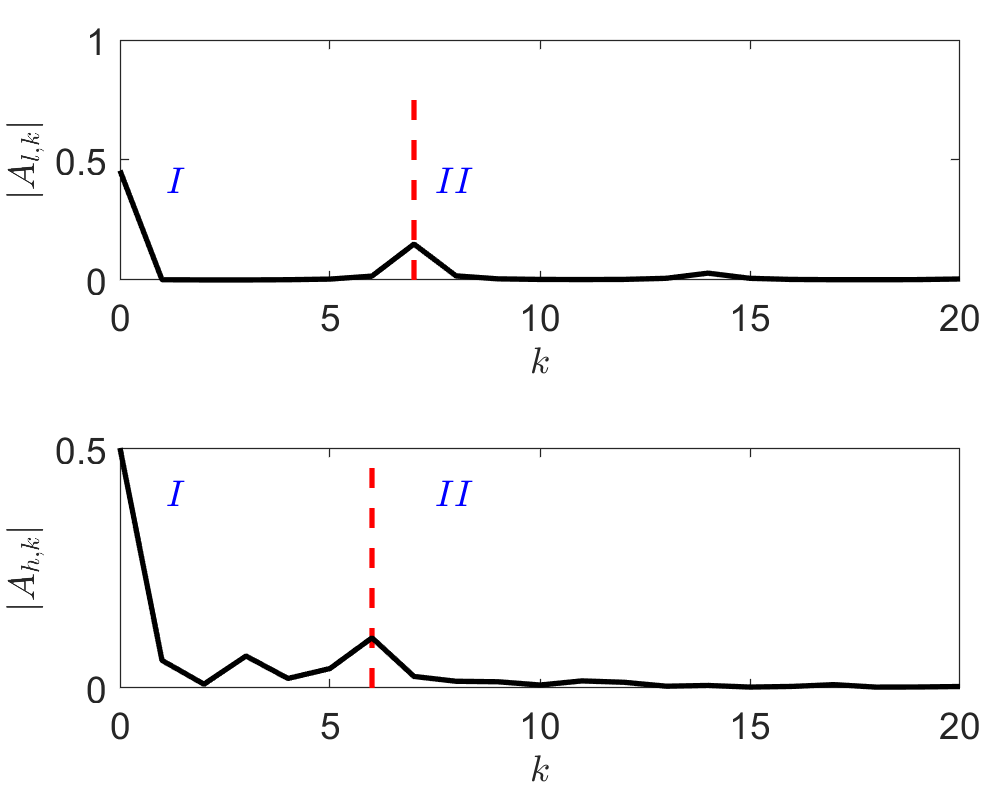}
\includegraphics[width=0.45\textwidth]{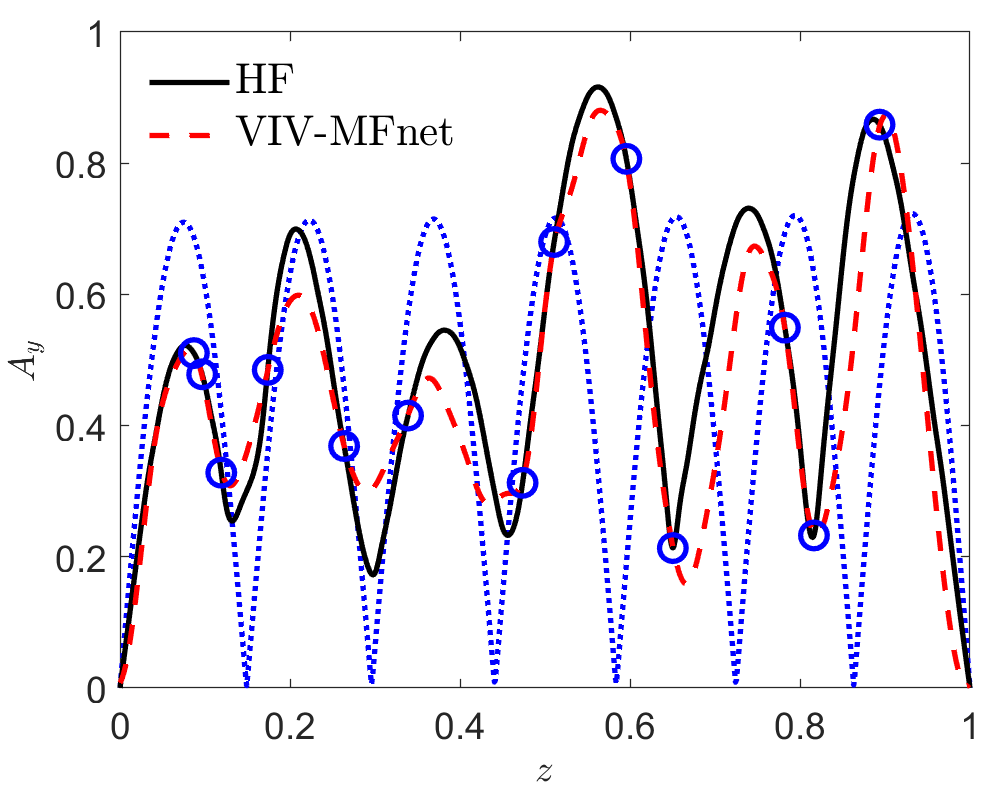}
}
\caption{\label{fig:mf_7}
Multi-fidelity predictions of displacements for uniform flow past a marine riser at $Ur=35.28$.
(a)  Left panel: Low- and high-fidelity data. LF: low-fidelity data from the VIVA model, HF: high-fidelity data from the MIT experiments. 
Right panel: Correlation between the low- and high-fidelity data.
(b) Left panel: Low- and high-fidelity data in the frequency space. First row: low-fidelity data, Second row: high-fidelity data. The critical and truncated wavenumbers are $k_c = 7$ and $k_t = 22$, respectively. 
Right panel: With $\lambda_I = 0.01$, and $\lambda_{II} = 0.25$. 
Blue dot: low-fidelity training data, Blue circle: high-fidelity training data.
}
\end{figure}

Note that the number of waves in all the cases tested above is the same for the data from VIVA simulations and experiments. However, there are some cases in which VIVA predicts a different number of waves from the experiments. Specifically, two different cases, i.e., $Ur = 17.23$, and 35.28, are considered. In both cases,  the number of half-wavelengths for the low-fidelity data is one more than that in the high-fidelity data. As for the first test case in Fig. \ref{fig:mf_3_4a}, the critical and truncated wavenumbers here are $k_c = 4$ and $k_t = 10$, respectively. We first employ 6 random samples as the training data. As illustrated in the last panel of Fig. \ref{fig:mf_3_4b}, the phase error between the low- and high-fidelity data can be corrected. In addition, the predicted maximum displacement is close to that from the experiment. To further improve the predicted accuracy, we then randomly add one more training data (blue cross in the right panel in Fig. \ref{fig:mf_3_4b}). We see that the error between the maximum displacements from the present method and experiment is further reduced.  In the second case (Fig. \ref{fig:mf_7}), the number of half-wavelengths in the low-fidelity data is 7, while we have 6 waves in the high-fidelity data. The critical and truncated wavenumbers are set as 7 and 22, respectively. The multi-fidelity predictions for both the phase as well as the amplitude are in good agreement with the experiments as 13 random training samples are used.

\begin{figure}[H]
\centering
\subfigure[]{\label{fig:shear_mf_3a}
\includegraphics[width=0.45\textwidth]{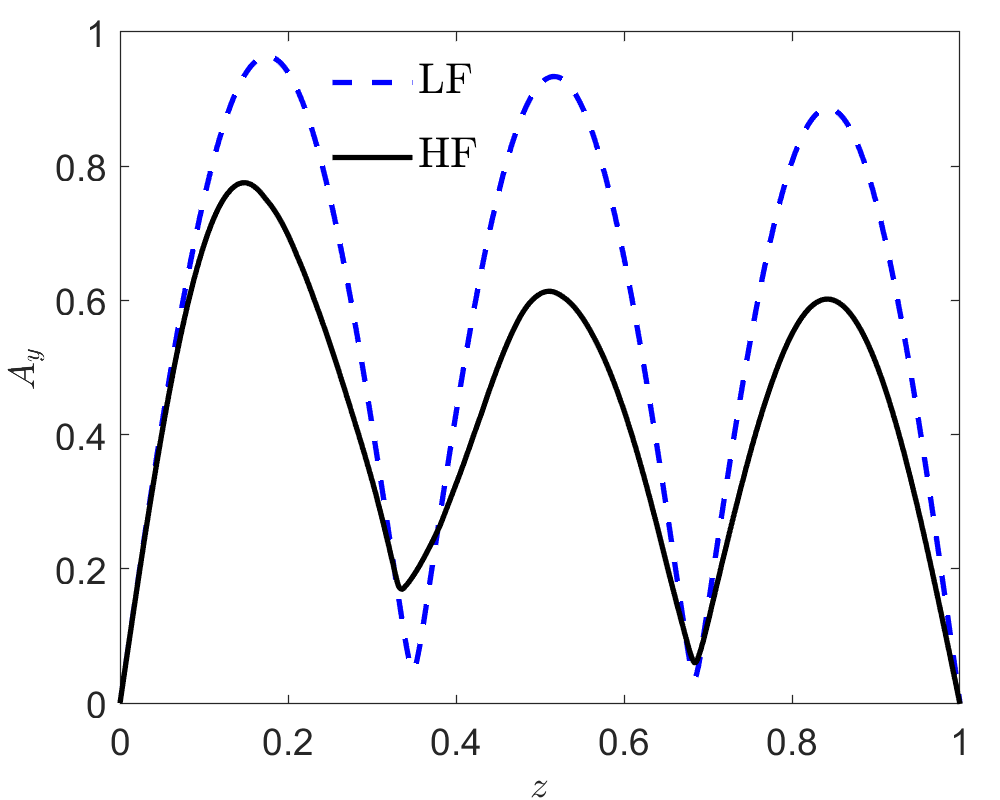}
\includegraphics[width=0.45\textwidth]{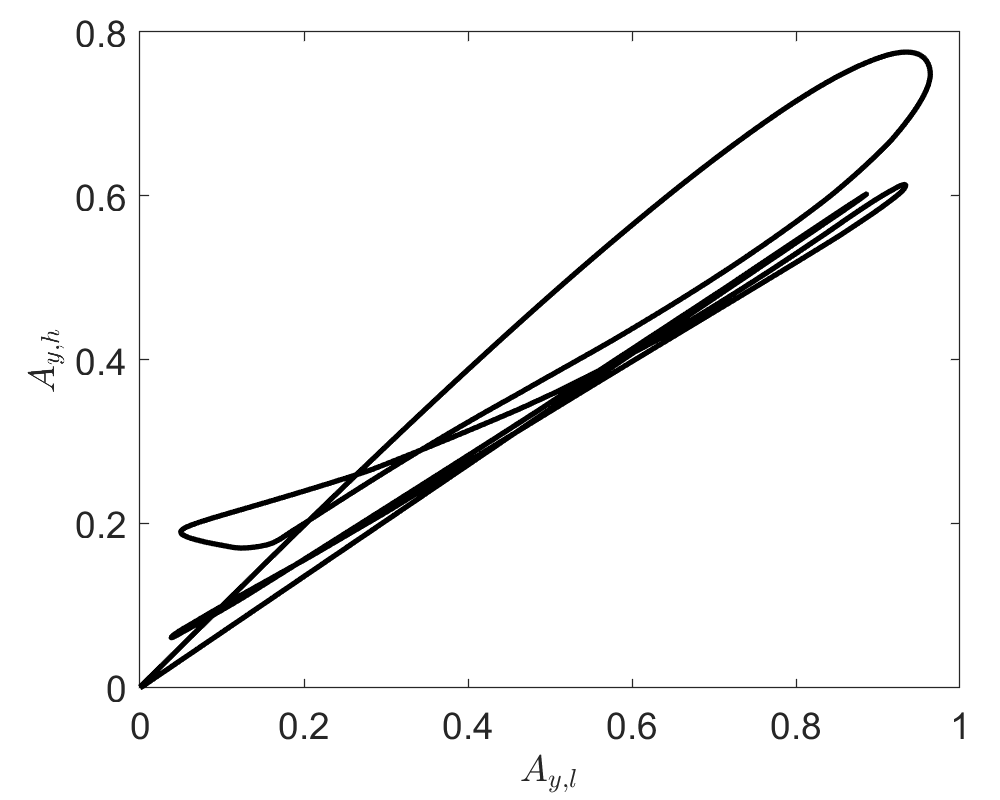}
}
\subfigure[]{\label{fig:shear_mf_3b}
\includegraphics[width=0.45\textwidth]{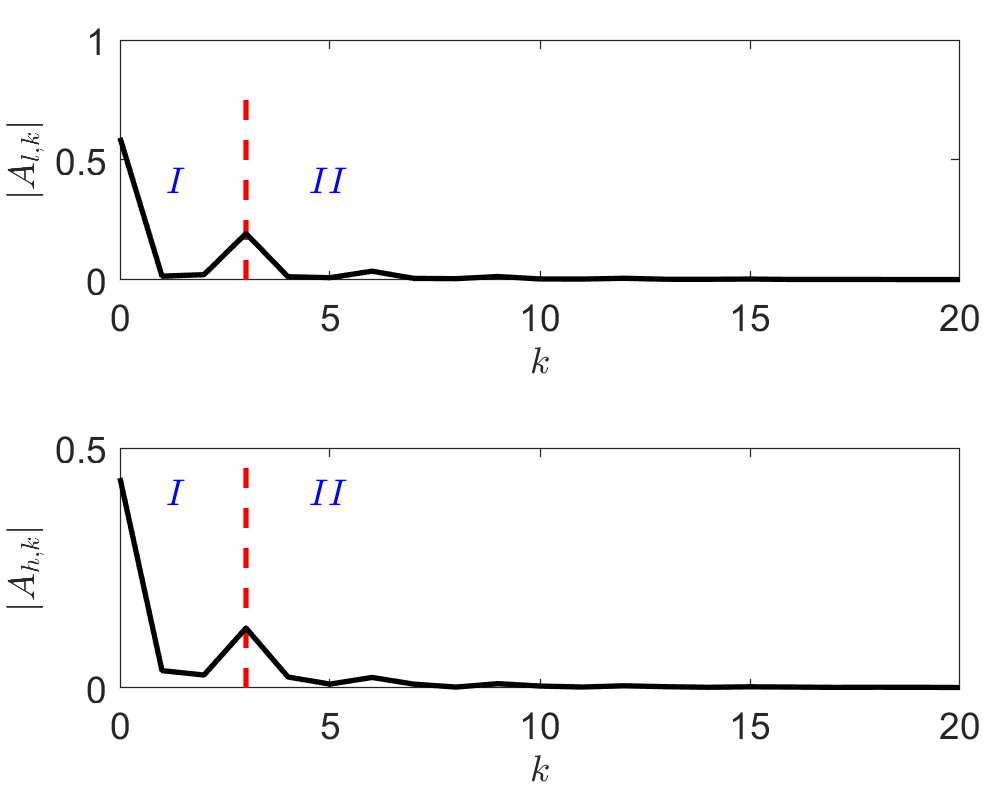}
\includegraphics[width=0.45\textwidth]{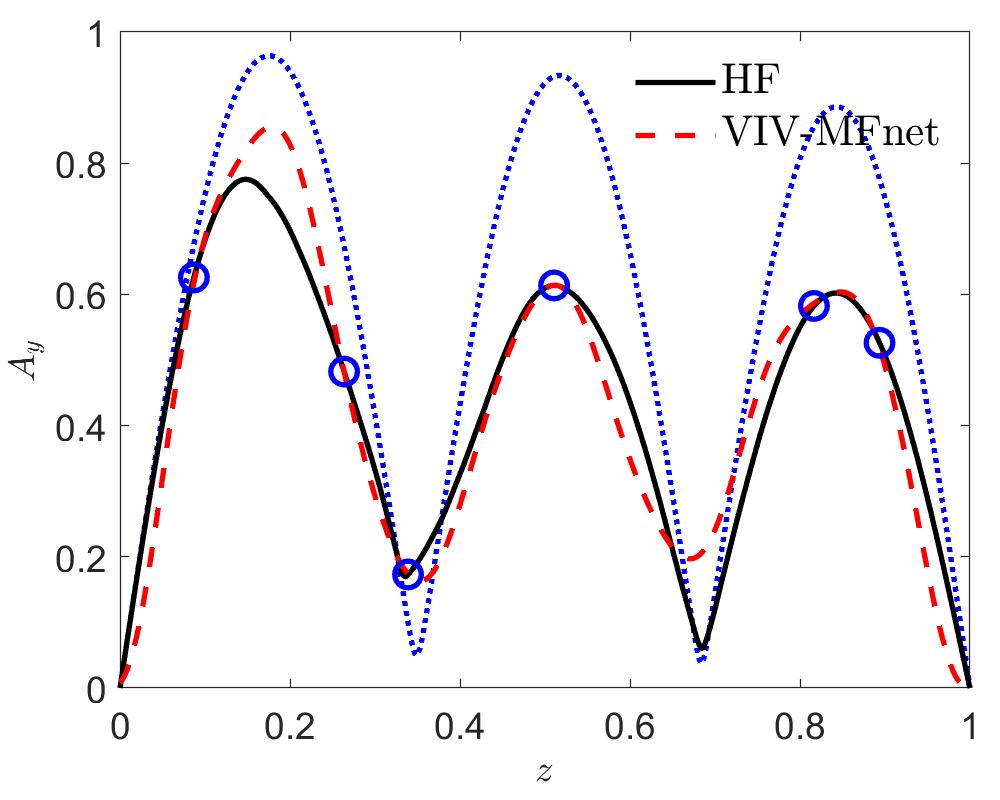}
}
\caption{\label{fig:shear_mf_3}
Multi-fidelity predictions of displacements for linearly sheared flow past a marine riser at $Ur=15.65$.
(a)  Left panel: Low- and high-fidelity data. LF: low-fidelity data from the VIVA model, HF: high-fidelity data from LES simulations. 
Right panel: Correlation between the low- and high-fidelity data.
(b) Left panel: Low- and high-fidelity data in the frequency space. First row: low-fidelity data, Second row: high-fidelity data. The critical and truncated wavenumbers are $k_c = 3$ and $k_t = 9$, respectively. 
Right panel: With $\lambda_I = 0.01$, and $\lambda_{II} = 0.25$. 
Blue dot: low-fidelity training data, Blue circle: high-fidelity training data.
}
\end{figure}

Next, We predict displacements of a riser caused by linear shear flow. Specifically,  the maximum velocity of the shear flow is $Ur = 15.65$. The low- and high-fidelity data are shown in Fig. \ref{fig:shear_mf_3}. Here the high-fidelity data are from a large eddy simulation (LES) rather than from experiments. Similarly, we assume that we have 6 random samples as the training data. The predictions from the multi-fidelity modeling agree well with results from the high-resolution LES.

\subsection{Results from UQ and active learning}
In this section, we use the Bayesian inference in Sec. \ref{sec:bayes} to quantify the predicted uncertainty. Moreover, we also conduct active learning to find the best location for adding more training samples based on the predicted uncertainty as in \cite{yang2020b}. Note that the uncertainty is represented by the standard deviation. 

\begin{figure}
\centering
\subfigure[]{\label{fig:bayes_mf_2a}
\includegraphics[width=0.3\textwidth]{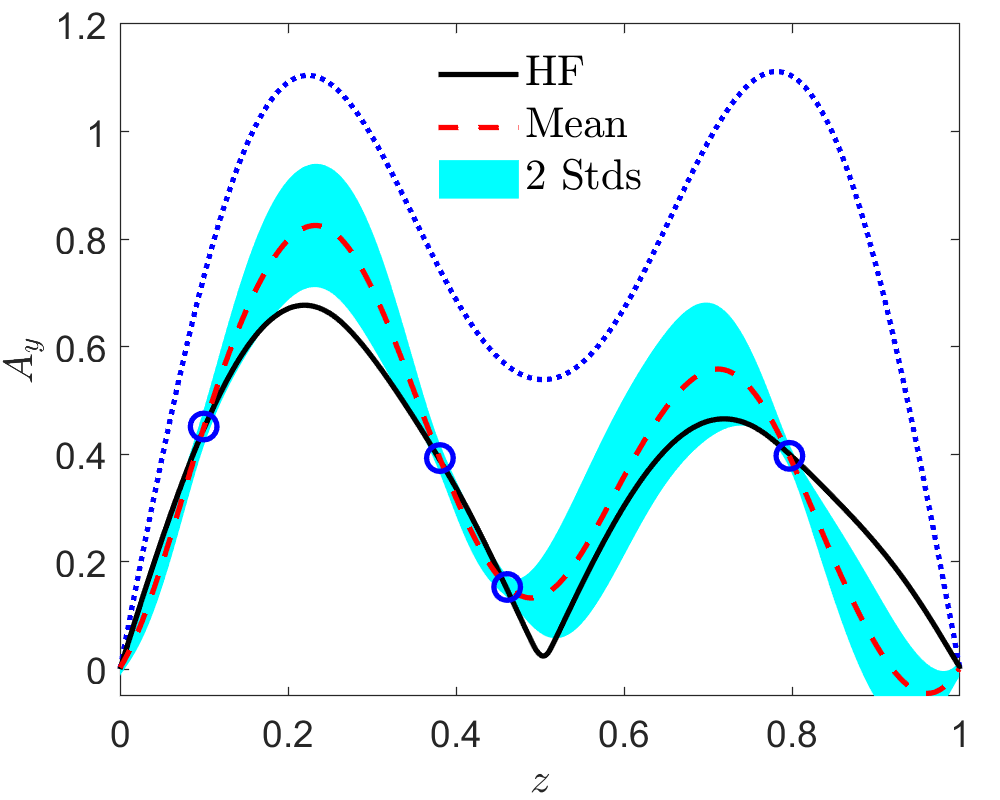}}
\subfigure[]{\label{fig:bayes_mf_2b}
\includegraphics[width=0.3\textwidth]{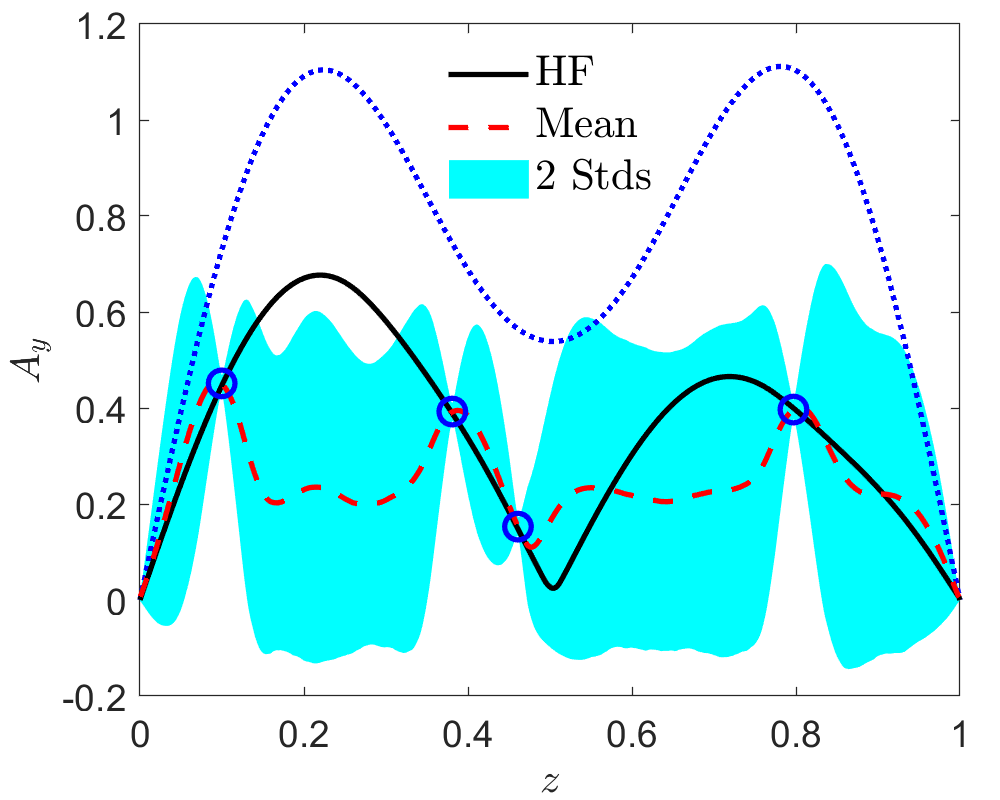}}
\caption{
Multi-fidelity Bayesian modeling of displacements for uniform flow past a marine riser at $Ur=12.66$: Importance of prior distributions.  
(a)  $P(\bm{\alpha}_I) \thicksim \mathcal{N}(0, 0.1^2)$, and $P(\bm{\alpha}_{II}) \thicksim \mathcal{N}(0, 0.01^2)$. 
(b)  $P(\bm{\alpha}_I) \thicksim \mathcal{N}(0, 0.1^2)$, and $P(\bm{\alpha}_{II}) \thicksim \mathcal{N}(0, 0.1^2)$. 
 HF: MIT experiment, Mean: predicted means from multi-fidelity predictions, 2 Stds: two standard deviations.
 Blue dot: low-fidelity training data, Blue circle: high-fidelity training data.
}
\label{fig:bayes_mf_2}
\end{figure}

We first test the performance of the BI using the case in Fig. \ref{fig:examplea}.  Note that both $k_c$ and $k_t$ for the present case are kept the same as in Fig. \ref{fig:mf_2a}, which are obtained from the low-fidelity data. In addition, we randomly sampled 4 high-fidelity data as the training data. As mentioned in Sec. \ref{sec:bayes}, $P({\bm{\alpha}_I}) \thicksim \mathcal{N}(0, 0.1^2)$, and $P({\bm{\alpha}_{II}}) \thicksim \mathcal{N}(0, 0.01^2)$ are employed as priors in the two subdomains. We observe in Fig. \ref{fig:bayes_mf_2} that: (1) the predicted means are in good agreement with the experimental data, and (2) the errors between the predicted means and the experiments are bounded by the two standard deviations in most regions. Furthermore, we also present the results with different prior distributions, i.e., $P({\bm{\alpha}_I}) \thicksim \mathcal{N}(0, 0.1^2)$, and $P({\bm{\alpha}_{II}}) \thicksim \mathcal{N}(0, 0.1^2)$. We see that: (1) the predicted means are not as accurate as in Fig. \ref{fig:bayes_mf_2a}, and (2) fluctuations are observed in the predicted uncertainties, especially for $z \in [0, 0.4]$, which may be attributed to the overfitting caused by the inappropriate prior distributions.  All these results demonstrate that the use of different prior distributions for different wavenumbers can enhance the predicted accuracy, which is consistent with the results in Sec. \ref{sec:map_result}.

\begin{figure}
\centering
\subfigure[]{\label{fig:bayes_mf_7a}
\includegraphics[width=0.3\textwidth]{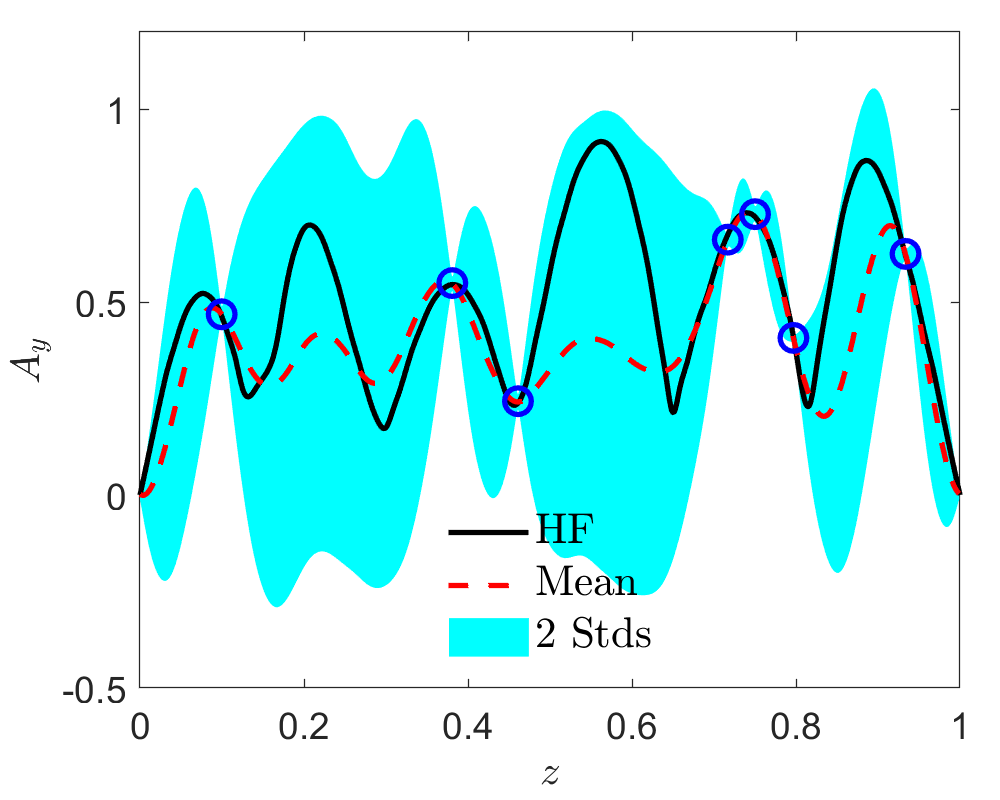}}
\subfigure[]{\label{fig:bayes_mf_7b}
\includegraphics[width=0.3\textwidth]{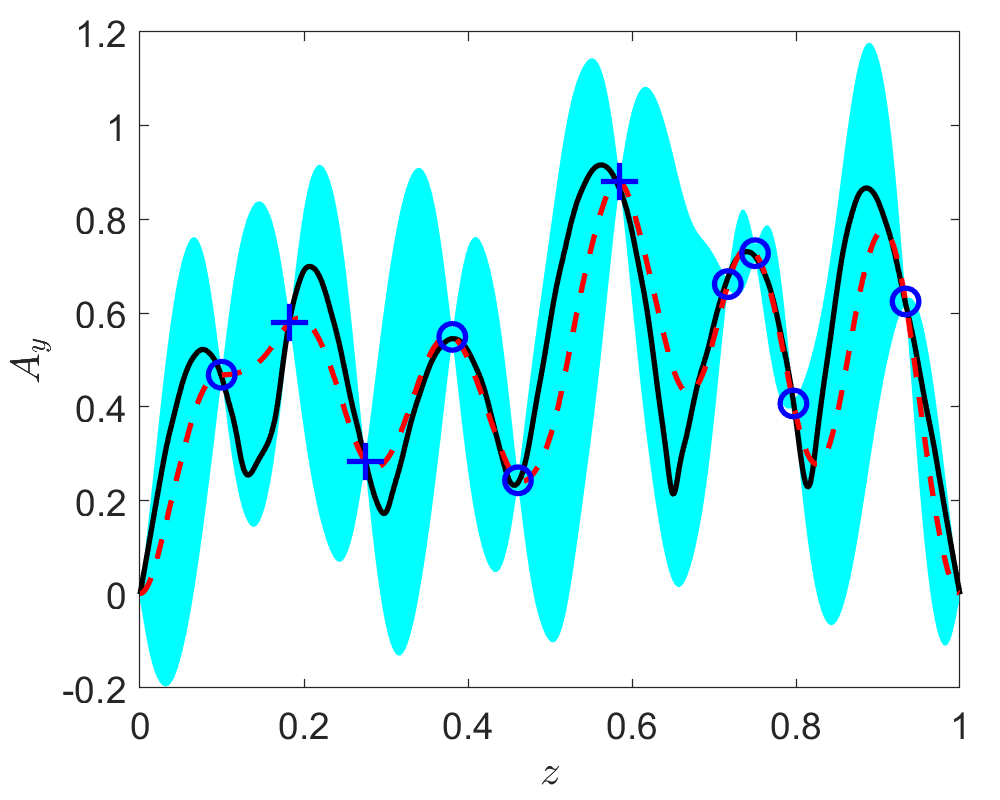}}
\subfigure[]{\label{fig:bayes_mf_7c}
\includegraphics[width=0.3\textwidth]{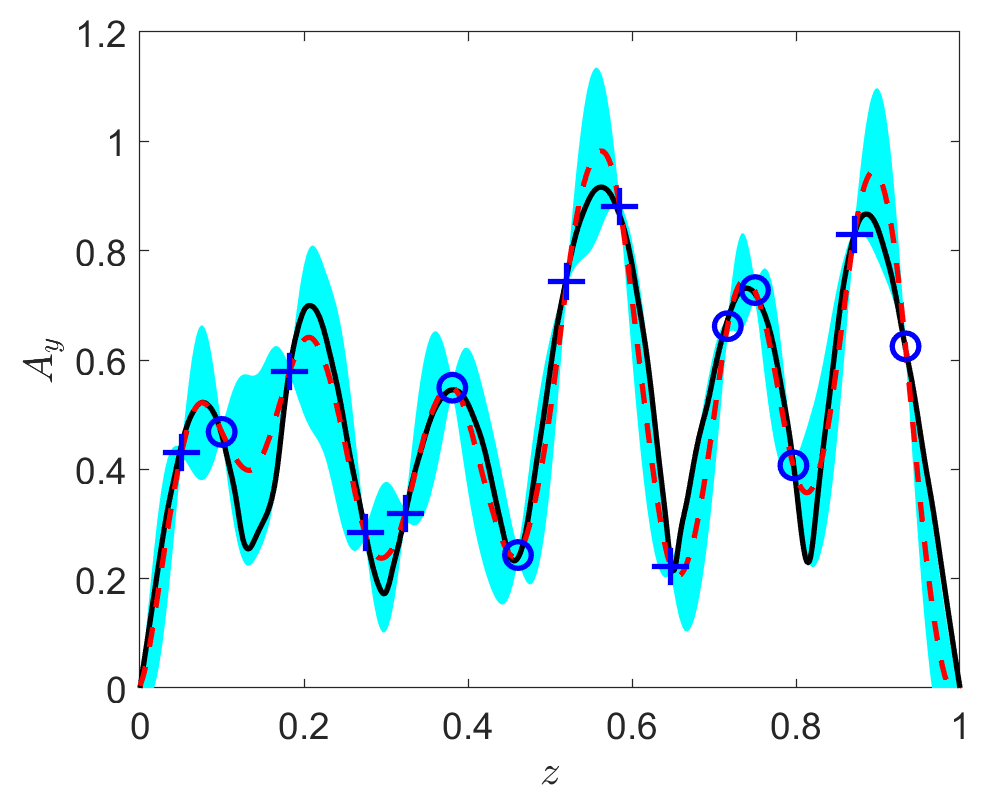}}
\caption{\label{fig:bayes_mf_7}
Multi-fidelity Bayesian modeling of displacements for uniform flow past a marine riser at $Ur=35.28$.
(a) With 7 training data. 
(b) With 10 training data.
(c) With 15 training data.
 Note that the low-fidelity training data are the same as in Fig. \ref{fig:mf_7}, which are not presented here. HF: MIT experiment, Mean: predicted means from multi-fidelity predictions, 2 Stds: two standard deviations. Blue plus: added high-fidelity training data based on the predicted uncertainty.
 }

\end{figure}

We now implement an active learning strategy based on the case in Fig. \ref{fig:mf_7}. Similarly, we employ  $P({\bm{\alpha}_I}) \thicksim \mathcal{N}(0, 0.1^2)$, and $P({\bm{\alpha}_{II}}) \thicksim \mathcal{N}(0, 0.01^2)$ as the prior distributions in the two subdomains. We start with 7 training data, which are randomly distributed in $ z \in [0, 1]$. As shown in Fig. \ref{fig:bayes_mf_7a}, the phase errors are well corrected while the predicted means do not fit the experiments quite well, which is also reflected in the large uncertainty. In addition, the computational errors between the predicted means and the experimental results are bounded by the two standard deviations. To reduce the uncertainty as well as enhance the predicted accuracy, we add more training data using the active learning strategy. Specifically, we can add one more training data at the location where the standard deviation is maximum, and then perform Bayesian inference using the new training dataset. We see in Figs. \ref{fig:bayes_mf_7b}-\ref{fig:bayes_mf_7c} that: (1) the predicted uncertainty decreases as we add more training data, (2) the predicted means are in good agreement with the experimental results as we employ 15 training data as shown in Fig. \ref{fig:bayes_mf_7c}, and (3) the computational errors are bounded by the two standard deviations for all results in Fig. \ref{fig:bayes_mf_7}.

\begin{figure}
\centering
\subfigure[]{\label{fig:bayes_mf_12a}
\includegraphics[width=0.3\textwidth]{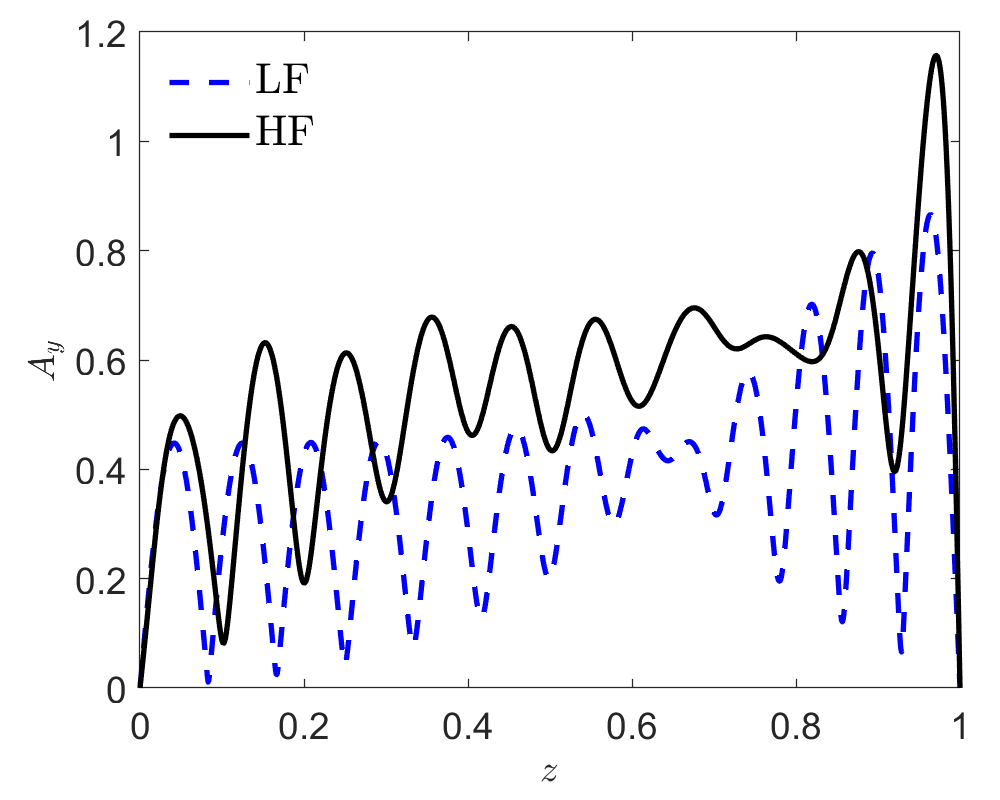}
\includegraphics[width=0.3\textwidth]{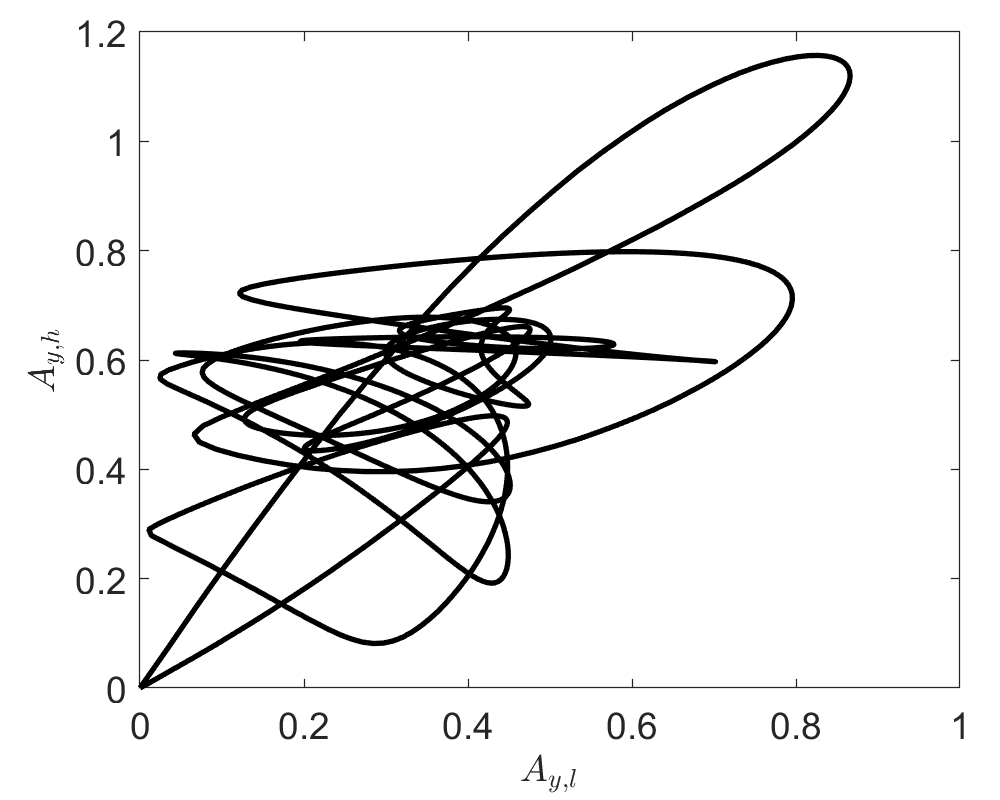}
\includegraphics[width=0.3\textwidth]{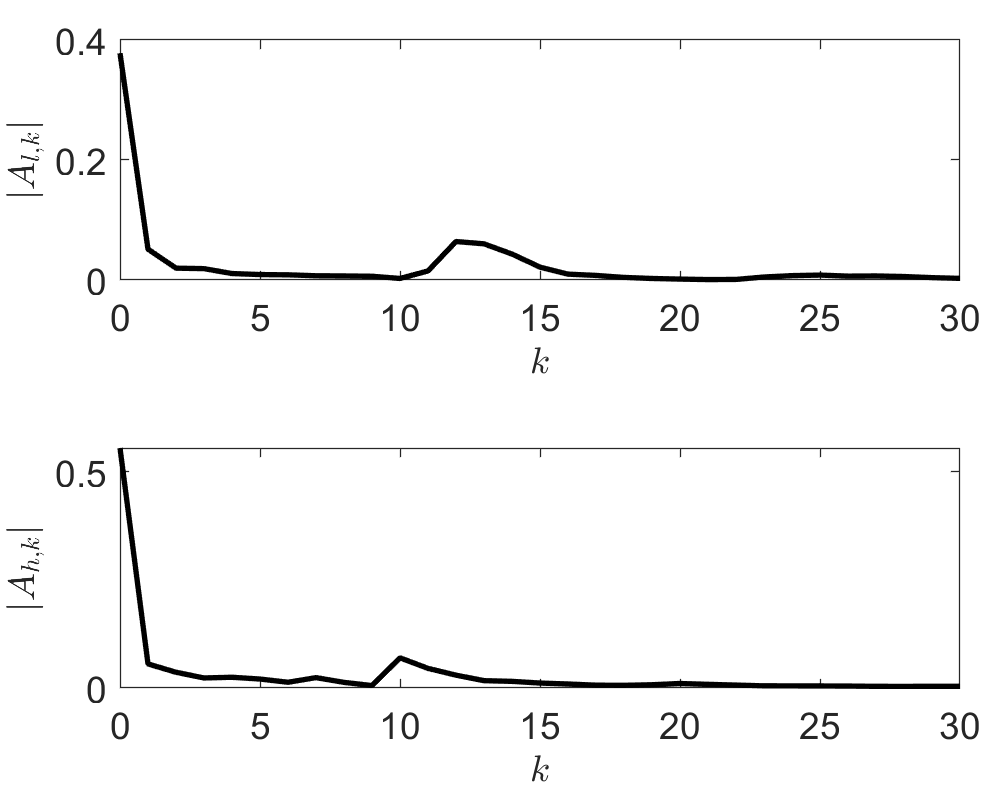}}
\subfigure[]{\label{fig:bayes_mf_12b}
\includegraphics[width=0.3\textwidth]{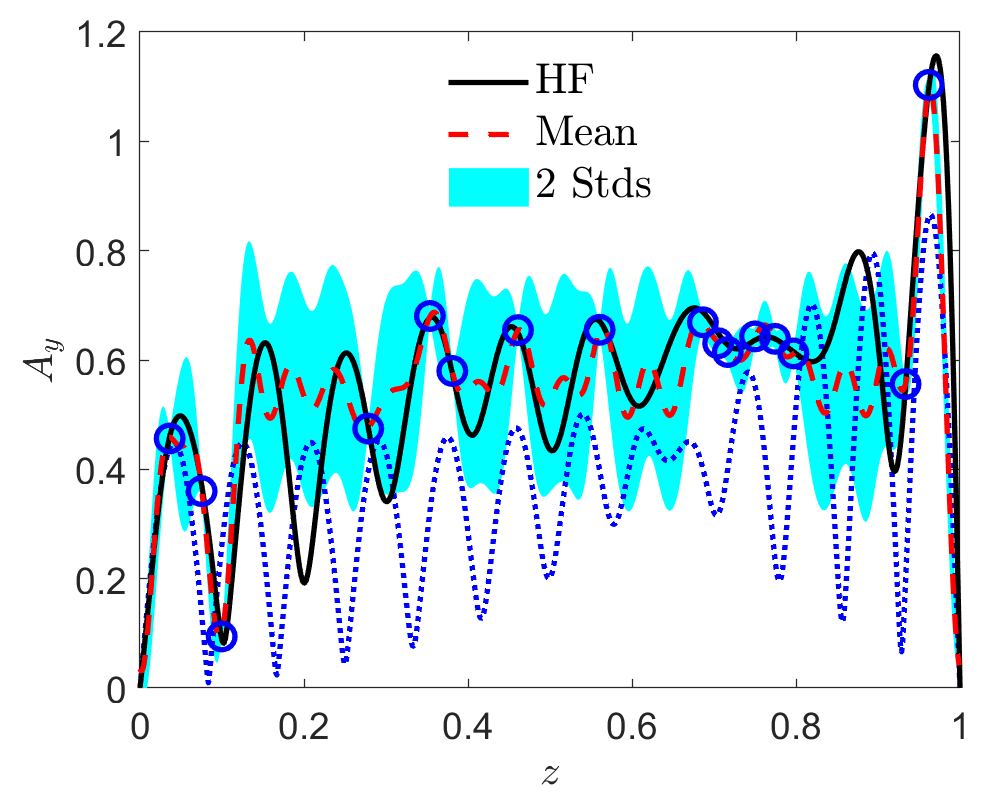}
\includegraphics[width=0.3\textwidth]{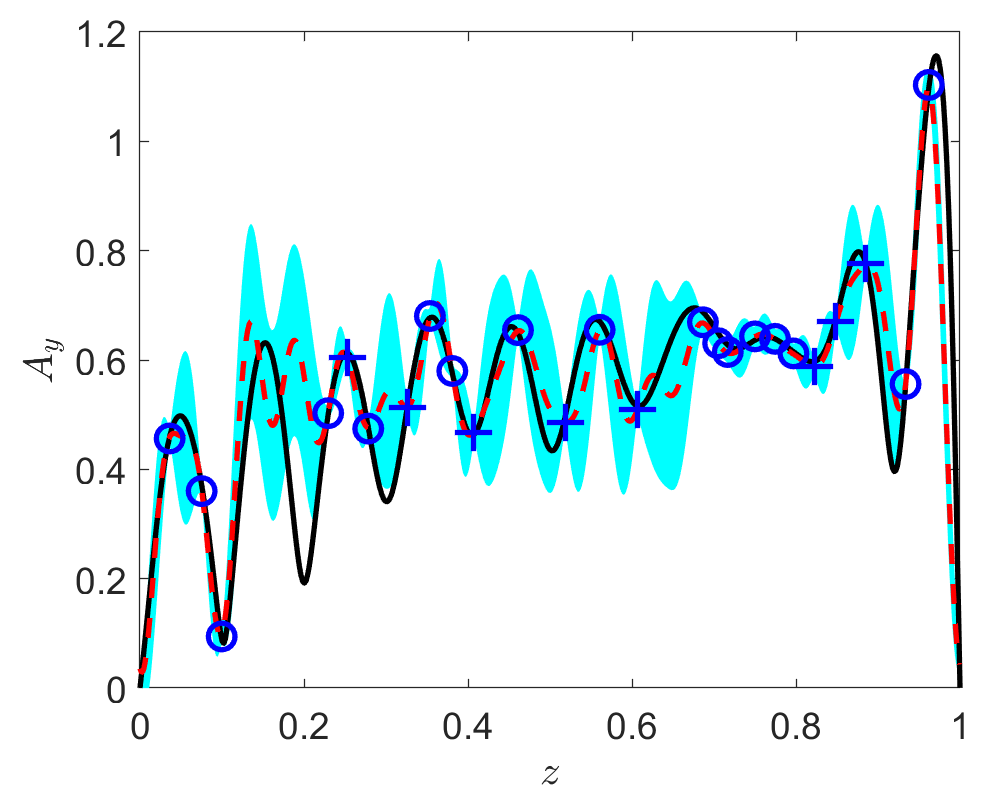}
\includegraphics[width=0.3\textwidth]{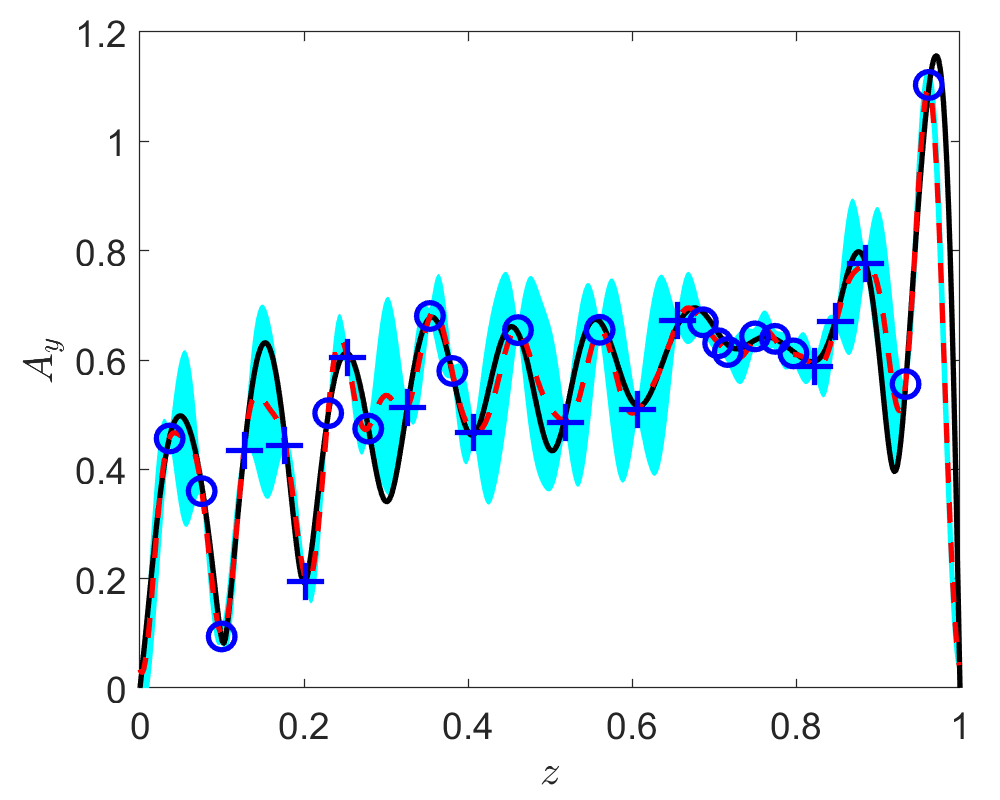}}
\caption{\label{fig:bayes_mf_12}
Multi-fidelity Bayesian modeling of displacements for linearly sheared flow past a marine riser (NDP case 2430).
(a) Left panel: Low- and high-fidelity data. LF: low-fidelity data that are from VIVA model, HF: high-fidelity data that are from the NDP data. 
Middle panel: Correlation between the low- and high-fidelity data.
Right panel: Low- and high-fidelity data in the frequency space. First row: low-fidelity data, Second row: high-fidelity data. The truncation wavenumber is $k_t = 20$.
(b) From left to right: 16, 24, and 28 training data. 
 HF: NDP data, Mean: predicted means from multi-fidelity predictions, 2 Stds: two standard deviations. 
 Blue dot: low-fidelity training data, Blue circle: initial high-fidelity training data,
 Blue plus: added high-fidelity training data based on the predicted uncertainty.
}

\end{figure}

Finally, we consider a shear flow case in which the low- and high-fidelity data are from VIVA simulations and the NDP experimental data, respectively (left panel in Fig. \ref{fig:bayes_mf_12a}). Based on the low-fidelity data in the modal space (right panel in Fig. \ref{fig:bayes_mf_12a}), modules for all the waves are below 0.1, we then use the same prior distribution for $\bm{\alpha}$ in the entire domain, i.e., $P(\bm{\alpha}) = \mathcal{N}(0, 0.02^2)$ (Note that 0.02 is about half of the maximum $A_{l,k}$). 
We start with 16 random training samples. As shown in Fig. \ref{fig:bayes_mf_12b}, the predicted means are not in good agreement with the NP data, which is also reflected in the large estimated uncertainty. Similarly, we can improve the predicted accuracy by performing active learning to add more training samples. In particular, the computational errors are bounded by the two standard deviations in most of the areas of $z \in [0, 1]$.

\section{Summary}
\label{sec:summary}
We present a multi-fidelity framework to predict vortex-induced vibrations of marine risers in modal space. This method, which we call VIV-MFnet, is capable of assimilating dense inexpensive low-fidelity data, obtained from the semi-empirical code VIVA, with a small set of high-fidelity data, consisting of displacement measurements obtained from either fully-resolved 3D LES or experiments on large-aspect-ratio flexible risers. Our aim is to enhance the predicted accuracy compared to single-fidelity modeling, which consists of performing regression with high-fidelity data only. In particular, the low-fidelity data are used to extract useful information on selecting basis functions as well as regularization in modal space to prevent overfitting, while the high-fidelity data are utilized to correct the phase/amplitude errors between the low- and the high-fidelity data. 

We applied two optimization approaches for the VIV-MFnet, the Maximum A Posteriori probability (MAP) estimate, and the Bayesian inference (BI). The selection of the two methods provides a balance between computational efficiency and advanced learning capability. Based on gradient descent, the first approach of MAP is more computationally efficient, about five times faster than BI, while the BI provides quantified uncertainty prediction, enabling active learning to adaptively improve the accuracy of prediction.

We apply the VIV-MFnet on multiple datasets from various sources: small-scale lab experiments, LES simulations, and large-scale experiments from the Norwegian Deepwater Program (NP), to test and demonstrate the effectiveness and robustness of the present method. Our main findings are as follows: 
\begin{enumerate}
    \item Using MAP, the vortex-induced vibration of a flexible cylinder in both uniform and shear flows is studied, and vibration displacements along the entire span are accurately reconstructed given only a small set of high-fidelity data and dense samples from low-fidelity data. 
    \item Comparison with the prediction results from state-of-the-art multi-fidelity deep neural networks shows that our approach outperforms these methods, which fail to capture the complex correlation between the low- and high-fidelity data that results in spatial phase errors.
    \item Using Bayesian inference, we are able to quantify the uncertainty in our predictions along the riser span, given limited high-fidelity observations.
    \item Guided by adaptive sampling strategy to minimize the global uncertainty, we illustrate that active learning can serve to adaptively select the optimal high-fidelity measurement locations along the span in order to improve the predicted accuracy. 
\end{enumerate}


Finally, it is noteworthy that the current approach using VIV-MFnet is computationally light, as it takes only about four minutes to learn the NDP experimental data of riser response on a personal laptop with 2-core CPU (Intel i7-8650U with 1.9 GHz). Such efficiency is important for future implementation of a digital twin of marine risers that is capable of providing real-time prediction and monitoring of production riser VIVs over their lefetime.

\section*{Acknowledgement}
We gratefully acknowledge support from the DigiMaR Consortium, consisting of ABS, ExxonMobil, Petrobras, SAIPEM, Shell, and Subsea 7.  G.E.K., Z.W., and X.M. would like to acknowledge the support by PhILMS grant DE-SC0019453, and OSD/AFOSR MURI grant FA9550-20-1-0358.

\bibliography{refs}

\begin{thebibliography}{10}
\expandafter\ifx\csname url\endcsname\relax
  \def\url#1{\texttt{#1}}\fi
\expandafter\ifx\csname urlprefix\endcsname\relax\def\urlprefix{URL }\fi
\expandafter\ifx\csname href\endcsname\relax
  \def\href#1#2{#2} \def\path#1{#1}\fi

\bibitem{wu2012review}
X.~Wu, F.~Ge, Y.~Hong, A review of recent studies on vortex-induced vibrations
  of long slender cylinders, Journal of Fluids and Structures 28 (2012)
  292--308.

\bibitem{williamson2004vortex}
C.~H.~K. Williamson, R.~N. Govardhan, Vortex-induced vibrations, Annual Review
  of Fluid Mechanics 36 (2004) 413--455.

\bibitem{wang2020review}
J.~Wang, D.~Fan, K.~Lin, A review on flow-induced vibration of offshore
  circular cylinders, Journal of Hydrodynamics 32~(3) (2020) 415--440.

\bibitem{gabbai2005overview}
R.~Gabbai, H.~Benaroya, An overview of modeling and experiments of
  vortex-induced vibration of circular cylinders, Journal of Sound and
  Vibration 282~(3-5) (2005) 575--616.

\bibitem{williamson2008brief}
C.~H.~K. Williamson, R.~N. Govardhan, A brief review of recent results in
  vortex-induced vibrations, Journal of Wind engineering and industrial
  Aerodynamics 96~(6-7) (2008) 713--735.

\bibitem{bearman2011circular}
P.~Bearman, Circular cylinder wakes and vortex-induced vibrations, Journal of
  Fluids and Structures 27~(5-6) (2011) 648--658.

\bibitem{bearman1984vortex}
P.~W. Bearman, Vortex shedding from oscillating bluff bodies, Annual Review of
  Fluid Mechanics 16~(1) (1984) 195--222.

\bibitem{dahl2007resonant}
J.~M. Dahl, F.~S. Hover, M.~S. Triantafyllou, et~al., Resonant vibrations of
  bluff bodies cause multivortex shedding and high frequency forces, Physical
  Review Letters 99~(14) (2007) 144503.

\bibitem{FanVIVJFM2019}
D.~Fan, Z.~Wang, M.~S. Triantafyllou, et~al., Mapping the properties of the
  vortex-induced vibrations of flexible cylinders in uniform oncoming flow,
  Journal of Fluid Mechanics 881 (2019) 815–858.

\bibitem{dahl2006two}
J.~M. Dahl, F.~S. Hover, M.~S. Triantafyllou, Two-degree-of-freedom
  vortex-induced vibrations using a force assisted apparatus, Journal of Fluids
  and Structures 22~(6) (2006) 807--818.

\bibitem{bourguet2011wake}
R.~Bourguet, Y.~Modarres-Sadeghi, G.~E. Karniadakis, et~al., Wake-body
  resonance of long flexible structures is dominated by counterclockwise
  orbits, Physical Review Letters 107~(13) (2011) 134502.

\bibitem{fan2017vortex}
D.~Fan, M.~S. Triantafyllou, Vortex induced vibration of riser with low span to
  diameter ratio buoyancy modules, in: The 27th International Ocean and Polar
  Engineering Conference, International Society of Offshore and Polar
  Engineers, 2017.

\bibitem{williamson1996vortex}
C.~H.~K. Williamson, Vortex dynamics in the cylinder wake, Annual Review of
  Fluid Mechanics 28~(1) (1996) 477--539.

\bibitem{sarpkaya2004critical}
T.~Sarpkaya, A critical review of the intrinsic nature of vortex-induced
  vibrations, Journal of Fluids and Structures 19~(4) (2004) 389--447.

\bibitem{chaplin2005laboratory}
J.~Chaplin, P.~Bearman, F.~H. Huarte, et~al., Laboratory measurements of
  vortex-induced vibrations of a vertical tension riser in a stepped current,
  Journal of Fluids and Structures 21~(1) (2005) 3--24.

\bibitem{chaplin2005blind}
J.~Chaplin, P.~Bearman, Y.~Cheng, et~al., Blind predictions of laboratory
  measurements of vortex-induced vibrations of a tension riser, Journal of
  Fluids and Structures 21~(1) (2005) 25--40.

\bibitem{braaten2004ndp}
H.~Braaten, H.~Lie, {NDP} riser high mode {VIV} tests, Norwegian Marine
  Technology Research Institute, Technical Report~(512394.00) (2004) 01.

\bibitem{modarres2010effect}
Y.~Modarres-Sadeghi, H.~Mukundan, J.~M. Dahl, et~al., The effect of higher
  harmonic forces on fatigue life of marine risers, Journal of Sound and
  Vibration 329~(1) (2010) 43--55.

\bibitem{modarres2011chaotic}
Y.~Modarres-Sadeghi, F.~Chasparis, M.~S. Triantafyllou, et~al., Chaotic
  response is a generic feature of vortex-induced vibrations of flexible
  risers, Journal of Sound and Vibration 330~(11) (2011) 2565--2579.

\bibitem{zheng2014fatigue}
H.~Zheng, R.~E. Price, Y.~Modarres-Sadeghi, et~al., On fatigue damage of long
  flexible cylinders due to the higher harmonic force components and chaotic
  vortex-induced vibrations, Ocean Engineering 88 (2014) 318--329.

\bibitem{vandiver2005high}
J.~Vandiver, H.~Marcollo, S.~Swithenbank, V.~Jhingran, et~al., High mode number
  vortex-induced vibration field experiments, in: Offshore Technology
  Conference, Offshore Technology Conference, 2005.

\bibitem{vandiver2006fatigue}
J.~K. Vandiver, S.~B. Swithenbank, V.~Jaiswal, V.~Jhingran, Fatigue damage from
  high mode number vortex-induced vibration, in: Proceedings of 25th
  International Conference on Ocean, Offshore and Arctic Engineering, 2006.

\bibitem{vandiver2009insights}
J.~K. Vandiver, V.~Jaiswal, V.~Jhingran, Insights on vortex-induced, traveling
  waves on long risers, Journal of Fluids and Structures 25~(4) (2009)
  641--653.

\bibitem{lucor2001vortex}
D.~Lucor, L.~Imas, G.~E. Karniadakis, Vortex dislocations and force
  distribution of long flexible cylinders subjected to sheared flows, Journal
  of Fluids and Structures 15~(3-4) (2001) 641--650.

\bibitem{bourguet2011vortex}
R.~Bourguet, G.~E. Karniadakis, M.~S. Triantafyllou, Vortex-induced vibrations
  of a long flexible cylinder in shear flow, Journal of Fluid Mechanics 677
  (2011) 342--382.

\bibitem{bourguet2013distributed}
R.~Bourguet, G.~E. Karniadakis, M.~S. Triantafyllou, Distributed lock-in drives
  broadband vortex-induced vibrations of a long flexible cylinder in shear
  flow, Journal of Fluid Mechanics 717 (2013) 361--375.

\bibitem{zhu2018wake}
H.~Zhu, D.~Zhou, Y.~Bao, R.~Wang, et~al., Wake characteristics of stationary
  catenary risers with different incoming flow directions, Ocean Engineering
  167 (2018) 142--155.

\bibitem{triantafyllou1999pragmatic}
M.~S. Triantafyllou, G.~Triantafyllou, Y.~Tein, et~al., Pragmatic riser {VIV}
  analysis, in: Offshore technology conference, Offshore Technology Conference,
  1999.

\bibitem{larsen2001vivana}
C.~M. Larsen, K.~Vikestad, R.~Yttervik, et~al., {VIVANA} theory manual,
  Marintek, Trondheim, Norway (2001).

\bibitem{roveri2001slenderex}
F.~E. Roveri, J.~K. Vandiver, Slenderex: Using shear7 for assessment of fatigue
  damage caused by current induced vibrations, in: Proceedings of the 20th
  International Conference on Offshore Mechanics and Arctic Engineering, 2001,
  pp. 3--8.

\bibitem{wang2020large}
Z.~Wang, D.~Fan, M.~S. Triantafyllou, et~al., A large-eddy simulation study on
  the similarity between free vibrations of a flexible cylinder and forced
  vibrations of a rigid cylinder, Journal of Fluid and Structures (under
  review).

\bibitem{gopalkrishnan1993vortex}
R.~Gopalkrishnan, Vortex-induced forces on oscillating bluff cylinders, Ph.D.
  thesis, Massachusetts Institute of Technology (1993).

\bibitem{fan2019robotic}
D.~Fan, G.~Jodin, T.~R. Consi, et~al., A robotic intelligent towing tank for
  learning complex fluid-structure dynamics, Science Robotics 4~(36) (2019).

\bibitem{xu2013experimental}
Y.~Xu, S.~Fu, Y.~Chen, et~al., Experimental investigation on vortex induced
  forces of oscillating cylinder at high {R}eynolds number, Ocean Systems
  Engineering 3~(3) (2013) 167--180.

\bibitem{chen2013hydrodynamic}
Y.~Chen, S.~Fu, Y.~Xu, et~al., Hydrodynamic characters of a near-wall circular
  cylinder oscillating in cross flow direction in steady current, Acta Physica
  Sinica 62~(6) (2013) 064701.

\bibitem{chang2011viv}
C.~C.~J. Chang, R.~A. Kumar, M.~M. Bernitsas, {VIV} and galloping of single
  circular cylinder with surface roughness at $3.0 \times 10^4< \mbox{Re} <1.2
  \times 10^5$, Ocean Engineering 38~(16) (2011) 1713--1732.

\bibitem{forrester2007multi}
A.~I. Forrester, A.~S{\'o}bester, A.~J. Keane, Multi-fidelity optimization via
  surrogate modelling, Proceedings of the Royal Society A: Mathematical,
  Physical and Engineering Sciences 463~(2088) (2007) 3251--3269.

\bibitem{perdikaris2017nonlinear}
P.~Perdikaris, M.~Raissi, A.~Damianou, et~al., Nonlinear information fusion
  algorithms for data-efficient multi-fidelity modelling, Proceedings of the
  Royal Society A: Mathematical, Physical and Engineering Sciences 473~(2198)
  (2017) 20160751.

\bibitem{bonfiglio2018multi}
L.~Bonfiglio, P.~Perdikaris, S.~Brizzolara, G.~Karniadakis, Multi-fidelity
  optimization of super-cavitating hydrofoils, Computer Methods in Applied
  Mechanics and Engineering 332 (2018) 63--85.

\bibitem{costabal2019multi}
F.~S. Costabal, P.~Perdikaris, E.~Kuhl, D.~E. Hurtado, Multi-fidelity
  classification using {G}aussian processes: accelerating the prediction of
  large-scale computational models, Computer Methods in Applied Mechanics and
  Engineering 357 (2019) 112602.

\bibitem{tian2020enhanced}
K.~Tian, Z.~Li, L.~Huang, K.~Du, L.~Jiang, B.~Wang, Enhanced variable-fidelity
  surrogate-based optimization framework by {G}aussian process regression and
  fuzzy clustering, Computer Methods in Applied Mechanics and Engineering 366
  (2020) 113045.

\bibitem{meng2020composite}
X.~Meng, G.~E. Karniadakis, A composite neural network that learns from
  multi-fidelity data: Application to function approximation and inverse {PDE}
  problems, Journal of Computational Physics 401 (2020) 109020.

\bibitem{zhang373multi}
X.~Zhang, F.~Xie, T.~Ji, Z.~Zhu, Y.~Zheng, Multi-fidelity deep neural network
  surrogate model for aerodynamic shape optimization, Computer Methods in
  Applied Mechanics and Engineering 373 (2020) 113485.

\bibitem{lee2019linking}
S.~Lee, F.~Dietrich, G.~E. Karniadakis, I.~G. Kevrekidis, Linking {G}aussian
  process regression with data-driven manifold embeddings for nonlinear data
  fusion, Interface focus 9~(3) (2019) 20180083.

\bibitem{neal2011mcmc}
R.~M. Neal, et~al., {MCMC} using {H}amiltonian dynamics, Handbook of {M}arkov
  {C}hain {M}onte {C}arlo 2~(11) (2011) 2.

\bibitem{betancourt2017conceptual}
M.~Betancourt, A conceptual introduction to {H}amiltonian {M}onte {C}arlo,
  arXiv preprint arXiv:1701.02434 (2017).

\bibitem{yang2020b}
L.~Yang, X.~Meng, G.~E. Karniadakis, B-{PINN}s: Bayesian physics-informed
  neural networks for forward and inverse {PDE} problems with noisy data,
  Journal of Computational Physics 415 (2021) 109913.

\end{thebibliography}

\end{document}